\documentclass[12pt,notoc]{JHEP3}

\usepackage{amsmath,amssymb,euscript,array,cite,mathrsfs}

\usepackage{epsfig}

\def\a{\alpha}
\def\b{\beta}
\def\c{\gamma}
\def\d{\delta}
\def\e{\epsilon}

\def\k{\kappa}
\def\l{\lambda}
\def\m{\mu}
\def\n{\nu}

\def\r{\rho}
\def\s{\sigma}
\def\t{\tau}
\def\u{\upsilon}
\def\w{\omega}

\def\D{\Delta}
\def\L{\Lambda}

\def\S{\Sigma}
\def\W{\Omega}

\def\vare{\varepsilon}

\def\det{{\rm det}}

\def\BcalF{\boldsymbol{\cal F}}

\def\BE{\boldsymbol{E}}
\def\Bvartheta{\boldsymbol{\vartheta}}
\def\Bone{\boldsymbol{1}}

\def\R{\boldsymbol{R}}

\def\BDelta{\boldsymbol{\Delta}}

\def\BA{\boldsymbol{A}}
\def\BB{\boldsymbol{B}}
\def\BC{\boldsymbol{C}}
\def\BK{\boldsymbol{K}}
\def\Bn{\boldsymbol{n}}

\def\Bh{\boldsymbol{h}}
\def\BC{\boldsymbol{C}}
\def\BOmega{\boldsymbol{\Omega}}
\def\orta{\overrightarrow}

\newcommand{\SLASH}[1]{{\raise.15ex\hbox{/}\mkern-12mu #1}}
\def\Dbarslash{\,\,{\raise.15ex\hbox{/}\mkern-12mu {\bar D}}}
\def\Dslash{\,\,{\raise.15ex\hbox{/}\mkern-12mu D}}
\def\delslash{\,\,{\raise.15ex\hbox{/}\mkern-9mu \partial}}
\def\delbarslash{\,\,{\raise.15ex\hbox{/}\mkern-9mu {\bar\partial}}}

\def\rta{\rightarrow}
\newcommand{\IM}{\operatorname{Im}}
\newcommand{\RE}{\operatorname{Re}}

\newcommand{\MAT}[1]{\begin{pmatrix} #1\end{pmatrix}}
\newcommand{\EQ}[1]{\begin{equation} #1 \end{equation}}

\newcommand{\SP}[1]{\begin{equation}\begin{split} #1
\end{split}\end{equation}}

\title{
The Refractive Index of Curved Spacetime II:\\
QED, Penrose Limits and Black Holes
}
\author{Timothy J. Hollowood, Graham M. Shore and Ross J. Stanley\\
Department of Physics,\\ Swansea University,\\
Swansea, SA2 8PP, UK.\\
E-mail: {\tt t.hollowood,g.m.shore,pyrs@swansea.ac.uk}}
\abstract{This work considers the way that quantum loop effects modify
  the propagation of light in curved space. The calculation of the
  refractive index for scalar QED is reviewed and then extended for
  the first time to QED with spinor particles in the loop. It is shown
  how, in both cases, the low frequency phase velocity can be greater
  than $c$, as found originally by Drummond and Hathrell, but
  causality is respected in the sense that retarded Green functions vanish
  outside the lightcone. A ``phenomenology'' of the refractive index
  is then presented for black holes, FRW universes and gravitational
  waves. In some cases, some of the polarization states propagate with
  a refractive index having a negative imaginary part indicating a potential
  breakdown of the optical theorem in curved space and possible instabilities.
}

\begin{document}

\section{Introduction}

The quantum theory of photon propagation in curved spacetime raises
many challenging questions, particularly in 
relation to the realization of causality and unitarity 
in quantum field theories with a fixed gravitational background. In
particular, in previous work
\cite{Hollowood:2007kt,Hollowood:2007ku,Hollowood:2008kq} we have
shown that vacuum 
polarization effects in curved spacetime 
lead to the phenomenon that curved spacetime acts like an optical
medium and can be described by means of a refractive index. However, 
this index has a novel analytic structure, 
compared with conventional optical
media, and standard formulae, notably the 
Kramers-Kronig dispersion relation are violated. This analytic
structure has an
origin in the geometry of null geodesic congruences and their
conjugate points which give rise 
to curved-spacetime-specific singularities in the Van Vleck-Morette
determinant and consequently the Green functions.

This line of research began with the apparent paradox of how to
reconcile the requirements of causality with
the observation by Drummond and Hathrell \cite{Drummond:1979pp} that at 
low frequencies the phase velocity of light in a gravitational background
can be superluminal ({\it i.e.}~the refractive index $n(\w) < 1$ for
$\w$ small). Since causality plays such an important r\^ole in the
theory, it is worthwhile stating exactly what is meant in the context
of a QFT. Causality in this context is often called 
``micro-causality'', and is the property that the
retarded/advanced Green functions only have non-vanishing support
inside (or on) the backward/forward lightcone. The requirement of
causality can be linked directly to the refractive index itself 
\cite{Hollowood:2008kq} leading to two conditions: (i) $n(\omega)\to1$
for large $|\omega|$; and (ii) $n(\omega)$ must be an analytic
function in the upper-half of the complex $\omega$ plane. The first condition
is equivalent to the requirement that the wavefront velocity, the high
frequency limit of the phase velocity, is equal to $c$
\cite{Shore:2003jx,Shore:2007um}. Notice that neither of these
conditions could be tested in the original work of Drummond and
Hathrell which was restricted to small $\omega$. However, prior to the
unlocking of the full frequency-dependence of $n(\omega)$ in
\cite{Hollowood:2008kq,Hollowood:2007ku,Hollowood:2007kt}, the
following argument led to a paradox.
In conventional dispersive, or dielectric, media 
the refractive index obeys the
familiar Kramers-Kronig dispersion relation 
\EQ{
n(\infty) ~~=~~ n(0) ~-~ \int_0^\infty{d\w\over\w}~\IM~n(\w) \ .
\label{aa}
}
Since $\IM n(\w)$ is constrained to be positive, being related by the
optical theorem (itself a consequence of unitarity) to a forward
scattering cross-section, eq.\eqref{aa} implies $n(\infty) < n(0)$.
This would mean a superluminal low frequency refractive index $n(0)<1$
would be incompatible with a causal limit $n(\infty)=1$.
The resolution, found in refs.\cite{Hollowood:2008kq,Hollowood:2007ku,Hollowood:2007kt}, is that spacetime behaves like an optical medium
for which the Kramers-Kronig relation no longer holds 
in the form \eqref{aa} and, furthermore,
$\IM n(\omega)$ is not constrained to be positive. 
One can trace this non-compliance to the 
novel analytic structure---compared with conventional dielectric media---of 
the refractive index of curved spacetime.

In curved spacetime, the refractive index will in general be a local quantity $n(x,\omega)$, since spacetime is not generally homogeneous, 
(as well as a matrix quantity in order to describe the
two polarization states of the photon). What we will find is that
$n(x,\omega)$ is consistent with {\it micro-causality\/},
as a consequence of analyticity in
the upper-half $\omega$ plane, and consequently one can write 
a version of the
Kramers-Kronig relation in the form
\EQ{
n(x,\infty) ~~=~~ n(x,0) ~-~ {1\over{i\pi}} {\cal P} 
\int_{-\infty}^\infty {d\w\over\w}~n(x,\w) \ ,
\label{ab}
}
which {\it does\/} hold. The usual relations that hold in models of dispersive
media, namely the fact that $n(\omega)$ is an even function of
$\omega$ and the property of {\it hermitian analyticity},
$n(\w - i\e) = n(\w + i\e)^*$, which together imply that 
$n(\omega+i\e)=n(-\omega+i\e)^*$, then lead to \eqref{aa}. 
Neither of these
relations hold in curved spacetime which, as we shall see, allows the
causality constraint $n(x,\infty)=1$
to hold even in the presence of a superluminal $n(x,0) < 1$.

Although this resolution of causality with a low-frequency superluminal
phase velocity is achieved through the modification of the Kramers-Kronig
relation to accommodate the novel analytic structure of $n(x,\w)$ in
curved spacetime, in the course of studying particular examples in
refs.\cite{Hollowood:2008kq,Hollowood:2007ku,Hollowood:2007kt}
we found cases where $\IM n(x,\w)$ is in fact negative. This is
unexpected since one would have expected this imaginary part to signal
the decay of the photon into real $e^+e^-$ pairs
and hence be strictly positive: it appears that the optical theorem
cannot hold---at least in its conventional form---in curved
spacetime \cite{Optical}. The correct
interpretation of a negative imaginary part, the fate of the optical
theorem and implication that the
photon modes have an increasing amplitude, will be left to future work;
one of the purposes of the present paper is to 
collect the ``phenomenological'' evidence on this issue by studying
several relevant spacetimes.

In order to establish these results, we calculated the complete frequency
dependence of the refractive index for scalar QED, obtaining a concise
formula involving the Van Vleck-Morette (VVM) matrix
\cite{Hollowood:2008kq}.\footnote{Usually one speaks of the Van
  Vleck-Morette determinant; here, we are interested in the actual
  matrix which we also call the ``VVM matrix''.}  
This makes clear the geometric origin of the
final result, and therefore its generality: loop corrections to 
Green functions in curved spacetime will generically have a richer
analytic structure than in flat spacetime.
The quantum corrections to photon propagation are governed by the 
one-loop vacuum polarization. As shown in refs.\cite{Hollowood:2007kt,
Hollowood:2007ku,Hollowood:2008kq}, both in a worldline formalism and
using the more conventional heat kernel (or proper time) formalism, when
the mass of the electron is much greater than the curvature scale and
so the ``geodesic approximation'' applies, this
is determined by a path integral over fluctuations around the
classical 
null geodesic traced by the photon; in turn, these are governed by the
geometry of geodesic deviation, explaining the origin of the VVM
matrix 
in the final result. The great simplification is that to leading order in 
a weak curvature expansion, the geometry of geodesic deviation is entirely
encoded in the Penrose limit \cite{Penrose} 
of the original spacetime around the 
photon's classical null trajectory. It follows that the whole problem
of determining the refractive index can be resolved simply by studying
the 
appropriate Penrose plane wave limit, with all the relevant information 
on the background spacetime being encoded in the plane wave profile
function $h_{ij}(u)$ to be defined in due course.

In this paper, we develop these fundamental ideas in a number of 
directions, in effect building up a phenomenology of the interesting and
counter-intuitive effects that can arise in what we have previously
termed ``quantum gravitational optics'' \cite{Shore:2003zc}.
We begin by reviewing and simplifying the conceptual basis of our 
calculation of the refractive index for scalar QED in curved
spacetime, highlighting the r\^ole played by the Penrose limit in
determining the 
vacuum polarization and the relation of the geometry of null geodesic
congruences to the analytic structure of the refractive index. 
Next we calculate the refractive index for spinor QED, introducing the 
necessary geometric tools, in particular the spinor parallel transporter
in a plane wave spacetime, and evaluating the vacuum polarization for
a massive spinor loop. The result is qualitatively similar to the simpler
case of scalar QED and displays the same generic analyticity properties,
but also shows some physically interesting differences in particular
models.

We then consider photon propagation in a variety of examples of
gravitational backgrounds in order to build up intuition about the
r\^ole 
of symmetries, singularities and time-dependence of the 
curved spacetime in the realization of causality and unitarity in QED. 
In each case, we need to find the Penrose limit corresponding to the 
classical photon trajectory of interest, then solve the geodesic 
deviation equations to construct the VVM matrix from which the  
vacuum polarization and refractive index follow. In most cases, we
have to resort to a numerical evaluation of the refractive index
itself.

The first examples are homogeneous plane waves. These are interesting
both as simple toy models and as the Penrose limits of certain geodesics 
in classical black hole and cosmological spacetimes near the
singularity. They have a large 
spacetime isometry group, an extension of the Heisenberg algebra, and
can be usefully realized as coset spaces. We consider two cases, the
symmetric plane waves originally studied in
refs.\cite{Hollowood:2008kq,Hollowood:2007ku,Hollowood:2007kt}
and the singular homogeneous plane  
waves which arise as near-singularity Penrose limits.
The next set of backgrounds that are considered are the black hole
spacetimes, especially 
the Schwarzschild and Kerr metrics. These are examples of Petrov type D
spacetimes and we can exploit the resulting simplifications to give a 
very general description of the Penrose limits for various choices of
geodesics. In particular, the Penrose limit corresponding to the  
principal null geodesics are flat, implying that the quantum corrections 
to the refractive index vanish identically, a result already observed
in the low-frequency limit. We also see how in general the Penrose
plane wave profile depends on the Walker-Penrose integral of motion 
\cite{Walker, Chandra} 
characterizing the classical trajectory and show how the
near-singularity limits reduce to singular homogeneous plane waves in
the Penrose limit. 

We then go on to consider FRW universes. In these spaces the definition of
the refractive index needs a slight modification because the
singularity in the past does not allow one to consider waves coming in
from past null infinity. Finally, we discuss the behaviour of the
refractive index in both a weak gravitational wave and a gravitational
shockwave. 

\section{Vacuum polarization and Photon Propagation in Curved Spacetime}

We begin by reviewing briefly the eikonal formalism and the derivation 
of the refractive index in the low-frequency effective action from the
QED effective action \cite{Hollowood:2008kq}.

In classical electrodynamics, the propagation of photons in curved 
spacetime is governed by the Maxwell equation $\nabla_\m F^{\m\n} = 0$,
with $F_{\m\n} = \nabla_\m A_\n - \nabla_\n A_\m$. In the eikonal
approximation, the electromagnetic field is written in terms of a
slowly-varying (with respect to the curvature scale) amplitude 
${\cal A}(x)$ and a rapidly-varying phase $\Theta(x)$ as follows:
\EQ{
A_\mu(x) = {\cal A}(x) \hat\vare_\m(x) e^{i\Theta(x)}\ ,
\label{ba}
}
where $\hat\vare_\m(x)$ is the polarization tensor. The wave-vector
is identified as $k_\m = \partial_\m\Theta$ and we fix the gauge
so that the two independent polarizations $\hat\vare_{(i)}$, 
$i=1,2$ satisfy the transverse condition $k\cdot\hat\vare_{(i)} = 0$, 
have vanishing component along $k^\m$ and are spacelike normalized 
such that $\hat\vare_{(i)}\cdot\hat\vare_{(j)} = \d_{ij}$.
The eikonal expansion is in powers of the frequency $\w$
and is valid in the regime $\w \gg\sqrt R$, with $R$ a typical curvature 
scale.\footnote{We can think of $R$ as the magnitude of a typical
  element of the Riemann tensor, with the understanding that 
  derivatives of it would count a factor of $\sqrt R$.} 
  The wave vector itself is ${\cal O}(\w)$ whereas ${\cal A}$ and 
$\hat\vare_\m$ are ${\cal O}(1)$, so at leading order, the Maxwell 
equation gives
\EQ{
k^2 = \partial\Theta \cdot \partial\Theta = 0\ .
\label{bb}
}
Since this also implies $k\cdot\nabla k^\m = 0$, it follows that the 
integral curves of the vector field $k^\m$ are null geodesics, which 
are identified as the classical trajectories of the photon.
At next-to-leading order, the eikonal equations describe the variation
of the amplitude and polarization along these null geodesics:
\SP{
k\cdot\nabla\,\hat\vare^\mu&=0 \ , \\
k\cdot\nabla\,\log{\cal A}&=-\tfrac12\nabla\cdot k \ .
\label{bc}
} 
The second of these relates the change in amplitude to the expansion
$\hat\theta \equiv \nabla_\m \hat k^\m$  
(with $\hat k^\m = \w^{-1}k^\m$), one of the optical scalars 
in the Raychaudhuri equations.

In order to study photon propagation in a general spacetime, it is
convenient to use a set of coordinates that are specially 
adapted to the vector field $k^\m$. These are the {\it Penrose
coordinates}, which are also known as 
``{\it adapted coordinates\/}'', $(u,V,Y^a)$, 
$a=1,2$. Here, $u$ is the affine parameter along a null geodesic, 
$V$ is the associated null coordinate and $Y^a$ are two orthogonal
spacelike coordinates. This choice corresponds to the embedding of 
the preferred null geodesic $\c$ (with $V=Y^a=0$), representing the 
classical photon trajectory, in a twist-free null congruence 
labelled by constant $V, Y^a$.
The metric $g_{\m\n}$ can always be written in terms
of these adapted coordinates in the form \cite{Blau2}:\footnote{In
  this paper, we work with a mostly plus signature for the metric.}
\EQ{
ds^2=-2du\, d V+C(u, V,Y^a)dV^2+2C_a(u, V,Y^b)
dY^a\,d V +C_{ab}(u, V,Y^c)dY^a\,dY^b\ .
\label{bd}
}
The eikonal phase is then taken to be $\Theta = -\w V$, the Fourier
mode appropriate for a metric with an isometry characterized by
a Killing vector $\partial_V$, so that $\hat k^\mu = (1,0,0,0)$.  
In terms of these coordinates the transport equations 
\eqref{bc} are
\SP{
\nabla_u \hat\vare^\mu = 0 \ , \\
\partial_u \log {\cal A} = -\tfrac12 \hat\theta \ ,
\label{be}
}
where the expansion scalar is
\EQ{
\hat\theta = \tfrac12 C^{ab}\partial_u C_{ab} = 
\partial_u \log\sqrt{g}\ ,
\label{bf}
}
where $C^{ab}$ denotes the inverse of $C_{ab}$ and
$g= -\det g_{\m\n}$.

Going beyond the classical theory and including the effects of 
vacuum polarization, the Maxwell equation is replaced by
\EQ{
\nabla_\n F^\n{}_\m ~=~
4\int \sqrt{g(x')}\,d^4x'\,\Pi^\text{1-loop}_{\m\n}(x,x')
A^\n(x')\ .
\label{bg}
}
where $\Pi_{\m\n}^\text{1-loop}(x,x')$ is the vacuum polarization 
tensor appearing in the one-loop effective action
\EQ{
\Gamma^\text{1-loop} ~=~ - \int
\sqrt{g(x)}\,d^4x\,\sqrt{g(x')}d^4x'\,A^\m(x)
\Pi^\text{1-loop}_{\m\n}(x,x')A^\n(x')
\ .
\label{bh}
}
In general, we find that the wave vector is no longer null and 
the physical light cone defined by $k^2$ no longer coincides with the
geometric null cones. In physical terms, the phase velocity of light 
is no longer necessarily $c$ and may be either sub- or super-luminal 
\cite{Drummond:1979pp}. It can also become polarization dependent,
{\it i.e.\/}~display gravitationally induced bi-refringence. 

To accommodate this, we modify the eikonal phase as follows:
\EQ{
\Theta=-\w\big(V\d_{ij}-\vartheta_{ij}(x;\w)\big)\ .
\label{bi}
}
where we now think of the phase as a $2\times 2$ matrix
with respect to the polarizations $\hat\vare_{(i)}$. 
With this form for the electromagnetic field $A_\m(x)$,
and using the metric \eqref{bd}, we find
\EQ{
\nabla_\m
F_{(i)}^{\m\n}=2\w^2\frac{\partial\vartheta_{ij}(x;\w)}
{\partial u}{\cal A}\hat\vare_{(j)}^\n e^{-i\w V}\ ,
\label{bj}
}
to leading order in the eikonal expansion. Provided 
$\vartheta_{ij}(x)$ is perturbatively small, 
{\it i.e.\/}~${\cal O}(\a)$, 
this corresponds to a refractive index matrix:
\EQ{
\Bn(x;\w)=\Bone+ \frac{\partial\Bvartheta(x;\w)}{\partial u}\ .
\label{bk}
}
The phase $\Bvartheta(x;\w)$ is then determined from the leading,
${\cal O}(\w^2)$ piece of the r.h.s.~of \eqref{bj} evaluated with
the eikonal ansatz \eqref{ba},\eqref{bi} for $A_\mu(x)$.
Putting all this together, we find the following compact expression
for the refractive index:
\EQ{
n_{ij}(x;\omega) ~=~ \d_{ij} - \frac{2}{\w^2} \int dx' \sqrt{g(x')}~
{{\cal A}(x')\over {\cal A}(x)}~
\hat\vare_{(i)}^\m(x) ~\Pi^\text{1-loop}_{\m\n}(x,x')~ 
\hat\vare_{(j)}^\n(x')~e^{i\w(V-V')} \ .
\label{bhh}
}
The polarizations which propagate with well-defined phase velocities
are the linear combinations of the basis $\hat\vare_{(i)}$ 
giving the eigenstates of $\Bn$.

To see how this works in the simplest case, we consider first the
low-frequency limit of the refractive index, which exhibits superluminal
phase velocities. This can be found by considering the modifications 
to the Maxwell equation following from the leading terms in a
derivative expansion of the one-loop effective action and was the
approach taken in the original work of Drummond and Hathrell
\cite{Drummond:1979pp}.  
The relevant terms in the low-frequency effective action to one loop 
are
\SP{
\Gamma~~=~\int d^4x\,\sqrt{g}
\Big[&-\frac14 ZF_{\mu\nu}F^{\mu\nu}~+~
d\nabla_\mu F^{\mu\lambda}\nabla_\nu
F^\nu{}_\lambda \\
&+~aRF_{\mu\nu}F^{\mu\nu} ~+~ bR_{\mu\nu}
F^{\mu\lambda}F^\nu{}_\lambda
~+~ cR_{\mu\nu\lambda\rho}F^{\mu\nu}F^{\lambda\rho}~~\Big] \ ,
\label{bii}
}
where the coefficients for both scalar and spinor QED can be read off 
from \cite{Shore:2002gw} and are given in \cite{Hollowood:2008kq}.
We find
\EQ{
a = 0\ , ~~~~~~
b = -\frac{\alpha}{720\pi m^2}\ , ~~~~~~
c = -\frac{\alpha}{1440\pi m^2}\ , ~~~~~~
d = \frac{\alpha}{480\pi m^2} \ ,
\label{bl}
}
for scalar QED, while
\EQ{
a = -\frac{\alpha}{144\pi m^2}\ , ~~~~~~
b = \frac{13\alpha}{360\pi m^2}\ , ~~~~~~
c = -\frac{\alpha}{360\pi m^2}\ , ~~~~~~
d = -\frac{\alpha}{30\pi m^2} \ ,
\label{bm}
}
for spinor QED, reproducing the original Drummond-Hathrell effective 
action \cite{Drummond:1979pp}. 
$Z = 1 + \tfrac \alpha{6\pi}\log{m^2\over\m^2}$ is the one-loop
wave-function renormalization factor.

To find the refractive index, we first evaluate the vacuum polarization
tensor from \eqref{bii}. Since this effective action is local, 
$\Pi^\text{1-loop}(x,x')$ is proportional to $\d(x,x')$ and \eqref{bhh} simplifies immediately since, for example, the amplitude factors cancel 
and the polarizations are evaluated at the same point. This immediately gives the general result for the low-frequency limit of the refractive 
index \cite{Drummond:1979pp,Shore:2003zc}
\EQ{
n_{ij}^\text{spinor}(0)=\delta_{ij}-2b R_{uu}\delta_{ij}-8cR_{uiuj}\ .
\label{bn}
}
where $R_{uiuj} \equiv R_{u\l u\r}\hat\vare_{(i)}^\l
\hat\vare_{(j)}^\r$.
So for scalar QED we find
\EQ{
n_{ij}^\text{scalar}(0)=\delta_{ij}-\frac{\alpha}{360\pi m^2}
\big(R_{uu}\d_{ij}+2R_{uiuj}\big)\ ,
\label{bo}
}
while for spinor QED,
\EQ{
n_{ij}^\text{spinor}(0)=\delta_{ij}-\frac{\alpha}{180\pi m^2}
\big(13R_{uu}\d_{ij}-4R_{uiuj}\big)\ ,
\label{bp}
}

The generalization of the QED effective action to all orders in 
derivatives was found in ref.\cite{Shore:2002gw,Shore:2002gn} 
and can be used to extend the expression \eqref{bn} for the refractive
index $n_{ij}(0)$ to a full perturbative expansion of $n_{ij}(\w)$
in powers of $\w$. This effective action consists of ``$RFF$'' operators acted on by functions of the Laplacian, {\it viz.}
\SP{
\Gamma ~=~ &\int d^4x \sqrt{-g}~ \biggl[ 
-{1\over4}ZF_{\m\n}F^{\m\n} ~+~ 
\nabla_\m F^{\m\l}\orta{d_0}\nabla_\n F^\n{}_\l \\
&+~{1\over m^2}\Bigl(\orta{a_0} R F_{\m\n} F^{\m\n}~ 
+~\orta{b_0} R_{\m\n} F^{\m\l}F^\n{}_{\l}~
+~\orta{c_0} R_{\m\n\l\r}F^{\m\n}F^{\l\r} \Bigr) \\
&+~{1\over m^4}
\Bigl(\orta{a_1} R \nabla_\m F^{\m\l} \nabla_\n F^\n{}_{\l}~  
+~\orta{b_1} R_{\m\n} \nabla_\l F^{\l\m}\nabla_\r F^{\r\n} \\
&+~\orta{b_2} R_{\m\n} \nabla^\m F^{\l\r}\nabla^\n F_{\l\r}
~+~\orta{b_3} R_{\m\n} \nabla^\m \nabla^\l F_{\l\r} F^{\r\n} 
~+~\orta{c_1} R_{\m\n\l\r} \nabla_\s F^{\s\r}\nabla^\l F^{\m\n} 
~\Bigr)\biggr] 
\label{bbi}
}
In this formula, the $\orta{a_n}$, $\orta{b_n}$, 
$\orta{c_n}$ are known ``form factor'' functions of three operators, 
{\it i.e.}
\EQ{
\orta{a_n} \equiv a_n\Bigl({\nabla_{(1)}^2\over m^2}, 
{\nabla_{(2)}^2\over m^2}, {\nabla_{(3)}^2\over m^2}\Bigr) \ ,
\label{bbj}
}
where the first entry $\nabla_{(1)}^2/m^2$ acts on the first following term
(the curvature), etc. The form factors can be extracted from the
very general background field effective action originally computed by
Barvinsky {\it et al.} \cite{BGVZone} and are described in detail
in ref.\cite{Shore:2002gw,Shore:2002gn}. 

Extracting the refractive index from \eqref{bbi} to leading order
in the eikonal approximation involves a number of subtleties, which
are discussed in detail in ref.\cite{Shore:2002gn}. The result is
an expression of the form
\EQ{
n_{ij}(\w) ~=~ \d_{ij} ~+~ 
\delta_{ij}
B\Bigl({2i\w {\hat k}\cdot\nabla\over m^2}\Bigr)\frac{R_{uu}}{m^2}~+~
C\Bigl({2i\w {\hat k}\cdot\nabla\over m^2}\Bigr)\frac{R_{uiuj}}{m^2} \ ,
\label{bbk}
}
where the constant coefficients are replaced by functions of the
operator ${\hat k}\cdot\nabla$, which describes the variation of the
curvature tensors along the original null geodesic $\c$.
Since $B(x)$ and $C(x)$ are real functions it
follows immediately that even perturbatively
at ${\cal O}(\w)$, a non-constant curvature along $\c$ will give 
rise to an imaginary part of the refractive index, which can be 
positive or negative depending on the variation of the curvature along
the geodesic.

The expression above \eqref{bbk}, captures all the terms
which are linear in the curvature. In the present work, we will
calculate an expression for the refractive index whose scope is much
wider because its sums up all
powers of the curvature (as well as its derivatives). The only assumptions are
that the curvature is weak, in the sense that $R\ll m^2$, and the
eikonal approximation is valid $\omega\gg\sqrt R$.\footnote{Both these
  conditions can be dropped if the spacetime is a plane wave.}
Schematically we find 
\EQ{
n_{ij}(\omega)=\delta_{ij}+\frac {\alpha R}
{m^2}F_{ij}\left(\frac{\omega\sqrt R}{m^2}\right)\ ,
}
where $R$ is a generic curvature scale.\footnote{This counting 
includes powers and
derivatives of the Riemann tensor, {\it e.g.\/} $\nabla_\nu
R_{\nu\sigma\rho\lambda}$ counts as $R^{3/2}$.}

To find the full frequency dependence of the refractive index, 
therefore, we need to find the exact form of the vacuum polarization
$\Pi^\text{1-loop}_{\m\n}$ acting on the eikonal ansatz for the
electromagnetic field. Even the complete expansion of the effective 
action to all-orders in derivatives is not sufficient to capture the 
non-perturbative behaviour of $n_{ij}(\w)$ as a function of frequency 
\cite{Hollowood:2007ku,Hollowood:2007kt,Shore:2002gn}.
This was achieved in refs.\cite{Hollowood:2007ku,Hollowood:2007kt},
in a calculation based on the worldline approach to QFT,
and subsequently in \cite{Hollowood:2008kq} using the more conventional
heat kernel, or proper time, representation of the propagators.

The key insight in both these methods is that when $R\ll m^2$ we can
use the ``geodesic approximation'' for the the propagators of the
electron and positron in the loop. When the loop is coupled to an 
external photon the geodesic approximation leads to a simple picture:
the electron and positron follow the original geodesic of the photon
before annihilating back into the photon. Therefore the 
vacuum polarization is determined by the geometry 
of geodesic fluctuations about the photon's classical trajectory 
and hence to leading order in ``weak'' curvature, i.e.~${\cal O}(R/m^2)$, 
the refractive index is governed entirely by the Penrose limit of the
original curved spacetime. This is because the Penrose limit is a
truncation of the original spacetime metric which captures the tidal  
forces corresponding to geodesic deviation. Since the Penrose limit 
describes a plane wave metric, we find the remarkable simplification
that at weak curvature, the refractive index for any given spacetime
may be calculated simply by considering propagation in the associated 
plane wave background.

We explain in detail how this comes about in Section 4, where 
we re-cap the essential points of the derivation of the refractive
index for scalar QED \cite{Hollowood:2008kq}. First, we discuss
the most important aspects of the Penrose limit, plane wave
spacetimes, geodesic deviation and the Van Vleck-Morette determinant
from the perspective of photon propagation in curved spacetime.

\section{Penrose limit and Geometry of Plane Wave Spacetimes}

In this section, we collect the essential results about the geometry
of the Penrose limit \cite{Penrose} and plane waves which will be used 
in the QFT calculations later in the paper. For further details and discussion, see the reviews in refs.\cite{Hollowood:2008kq} 
and \cite{Blau2}.

\subsection{Geodesic deviation and the optical tensors}

Before introducing the Penrose limit, we describe the basic geometry 
of geodesic deviation in the context of the general metric \eqref{bd}
in coordinates adapted to the null congruence around $\c$.
We therefore consider the {\it connecting vector} $z^\m$ which 
connects corresponding points on neighbouring geodesics in
the congruence and study its evolution along $\c$. This is determined
by the requirement that the Lie derivative of $z^\m$ along $\c$
vanishes, {\it i.e.}
\EQ{
{\cal L}_{\hat k} z^\m ~\equiv~
\hat k \cdot \nabla z^\m - (\nabla_\n \hat k^\m) z^\n ~=~ 0 \ .
\label{ca}
}
This implies
\EQ{
\nabla_u z^\m ~=~ \W^\m{}_\n z^\n \ ,
\label{cb}
}
where $\W_{\m\n} = \nabla_\m \hat k_\n$, which plays the role
of a connection for the Lie derivative, is symmetric since 
the vector field $\hat k^\m$ is a gradient flow.

To describe geodesic flow and the Raychaudhuri equations, it is 
sufficient\footnote{Here, we follow the approach presented by 
Wald \cite{Wald:1984rg}. The 4-dim space ${\cal V}$ of tangent 
vectors $z^\m$ is first restricted to a 3-dim subspace 
$\tilde{\cal V}$ by the condition $\hat k\cdot z = 0$. The final 
2-dim vector space $\hat{\cal V}$ is identified as the vector space 
of equivalence classes of vectors in $\tilde{\cal V}$ with vectors differing only by the addition of a multiple of $\hat k^\m$ deemed equivalent. In terms of the Penrose (adapted) coordinate
system \eqref{bd}, it is clear that $\hat{\cal V}$ is realized
by restricting to vectors $z^\m$ whose only non-vanishing components
are $z^a$. Similarly for the fundamental tensor field $\W_{ab}$.
Note that $\W_{\m\n}$ automatically satisfies 
$\hat k^\m \W_{\m\n} = 0$ by virtue of the null geodesic equation 
$\hat k\cdot\nabla\hat k^\n = 0$.}
to consider connecting vectors with transverse components $z^a$ only.
In the Penrose coordinates \eqref{bd}, we have\footnote{
All the covariant equations in this sub-section are valid in an
arbitrary coordinate system. In Penrose coordinates, \eqref{cc}
is simply $\partial_u z^a = 0$. In particular, this confirms that 
$Y^a$ itself is a suitable connecting vector, with the null geodesics
given simply by $Y^a = {\rm constant}$.}
\EQ{
\nabla_u z^a ~=~ \W^a{}_b z^b \ ,
\label{cc}
}
with
\EQ{
\W_{ab} = \frac{1}{2} \partial_u C_{ab}\ ,
\label{cd}
}
It follows from \eqref{cc} that the transverse components of the 
connecting vector satisfy the geodesic deviation equation:
\EQ{
\nabla_u \nabla_u z^a ~=~ - R^a{}_{ubu} z^b ~\equiv~ -h^a{}_b z^b\ ,
\label{ce}
}
where\footnote{In Penrose coordinates,~
$h^a{}_b = -\partial_u \W^a{}_b -\W^a{}_c \W^c{}_b\ ,$~ 
or equivalently,~ 
$h_{ab} = -\partial_u \W_{ab} +\W_{ac} \W^c{}_b \ .$ }
\EQ{
h^a{}_b ~=~ -\nabla_u \W^a{}_b -\W^a{}_c \W^c{}_b \ ,\\
\label{cf}
}
that is,
\EQ{
R_{aubu} ~=~ -\frac{1}{2} \partial_u^2 C_{ab}
~+~ \frac{1}{4} \partial_u C_{ac}~ C^{cd}~ \partial_u C_{db} \ .
\label{cg}
}
A solution $z^a$ of \eqref{ce} is known as a ``Jacobi field'' on 
the null geodesic $\c$. 

The important point here is that the transverse geodesic 
fluctuations are controlled entirely by the metric components 
$C_{ab}$ in \eqref{bd}. As we see below, this underlies the key r\^ole 
to be played by the Penrose limit.

The nature of the geodesic flow is most elegantly summarized in the
Raychaudhuri equations for the optical tensors. These are defined
from $\W_{ab}$ as:
\EQ{
\Omega_{ab} ~=~ \frac{1}{2} \hat\theta C_{ab} + \hat\s_{ab} + 
\hat\w_{ab} \ ,
\label{ch}
}
where $\hat\theta$, $\hat\s_{ab}$, $\hat\w_{ab}$ are respectively the 
expansion, shear and twist of the null congruence. Here, the twist
is vanishing by construction, since $\hat k^\m$ is by definition a 
gradient field, and for this reason \eqref{bd} embeds the geodesic
$\c$ in a twist-free null congruence. 

The Raychaudhuri equations are equivalent to \eqref{cf},
which reads
\EQ{
\nabla_u \W_{ab} ~=~ -\W_{ac} C^{cd} \W_{db} ~-~ R_{aubu} \ .
\label{ci}
}
In terms of the optical scalars:
\SP{
\partial_u \hat\theta ~&=~ -\tfrac12 \hat\theta^2 - 
\hat\s_{ab} \hat\s^{ab} - R_{uu} \ , \\
\nabla_u \hat\s_{ab} ~&=~ -\hat\theta \hat\s_{ab} - {\cal C}_{aubu} \,
\label{cj}
}
where the Weyl tensor ${\cal C}_{aubu}$ is the trace-free part
of the Riemann tensor \eqref{cg}. 

Notice that the equation for the expansion immediately shows that
$\partial_u \hat\theta \leq 0$ everywhere along the geodesic
provided the null energy condition $R_{uu} > 0$ holds, implying
a generic focusing of the null congruence.
Moreover, if we define the shear scalar as $\hat\s = \sqrt{2}
\hat\s_{ab} \hat\s^{ab}$ and introduce Newman-Penrose scalars
$\Phi_{00} = \tfrac12 R_{uu}$ and (in a null basis $m^\m, \bar m^\m$ 
for the transverse directions aligned with the eigenvalues of 
$\s_{ab}$) $\Psi_0 = {\cal C}_{umum}$,  
we can write the Raychaudhuri equations as
\EQ{
\partial_u (\hat\theta \pm \hat\s) ~=~
-\tfrac12 (\hat\theta \pm \hat\s)^2 - 2(\Phi_{00} \mp \Psi_0)\ .
\label{ck}
}
Physically, this shows that at a given point along the geodesic $\c$,
the congruence must be focusing in at least one of the transverse 
directions. While locally, focus/focus and focus/defocus are permitted, 
the null energy condition prohibits a congruence with defocus/defocus.

This illustrates a crucial theorem \cite{Wald:1984rg}, that 
for a spacetime satisfying the ``null generic condition'' 
($\Phi_{00} \neq 0$ or $\Psi_0 \neq 0$ at some point) and the
null energy condition, every complete null geodesic possesses
a pair of {\it conjugate points}. Two points $p$ and $q$ on a null
geodesic are said to be conjugate if there exists a Jacobi field 
$z^a \in \hat{\cal V}$ 
which is not identically zero but which vanishes at both
$p$ and $q$. Loosely speaking, this implies there is an
``infinitesimally deformed'' null geodesic intersecting $\c$ at 
both $p$ and $q$. As we see in the next section, this is precisely
the property of geodesic fluctuations which is reflected in the 
quantum field theory. Notice though that this definition of
conjugate points is linked to infinitesimal deviations and does 
not necessarily lift to geodesics of the full metric; in particular,
the existence of conjugate points $p$ and $q$ does not imply that 
there is an actual geodesic, other than $\c$, joining $p$ and $q$.

\vskip0.2cm
Another important set of coordinates, specifically designed to study
fluctuations around a given null geodesic, are {\it null Fermi 
coordinates.}  Fermi normal coordinates are analogues of the familiar 
Riemann normal coordinates, in which the connection coefficients
vanish locally, extended from a given point to the whole of a
given geodesic. The formalism for constructing Fermi coordinates 
appropriate for a null geodesic was developed in 
ref.\cite{Blau:2006ar} and we refer to this paper for further details.

To construct a null Fermi coordinate system around $\c$, we begin 
by introducing a pseudo-orthonormal frame $E^A = E^A{}_\m~ dx^\m$ 
($A = u,v,i=1,2$) such that
\EQ{
ds^2\big|_\c ~=~ -2 E^u E^v +\d_{ij} E^i E^j \ ,
\label{cl}
}
which is parallel-transported along $\c$ and where $E_u{}^\m$ is 
the tangent vector. The Fermi coordinates $x^A = (u,v,y^i)$ are 
then essentially the coordinates along the directions defined by this 
frame, with $u$ playing the role of affine parameter. Precisely, 
we define
\EQ{
x^A ~=~ (u, x^\a = s E^\a{}_\m(u_0) \dot x^\m(0)) \ , ~~~~~~
(\a = v,i=1,2)\ ,
\label{cm}
}
where $x^\m(s)$ are geodesics emanating from a point $u_0$ on $\c$;
equivalently, 
\EQ{
{\partial x^A \over \partial x^\m}\bigg|_\c ~=~ E^A{}_\m \ .
\label{cn}
}
Just as with Riemann normal coordinates, the key property of Fermi
coordinates is that to linear order in an expansion in powers
of the ``transverse'' coordinates $x^\a$ around $\c$, the connection
coefficients vanish, 
$\Gamma^A_{BC}\big|_\c = {\cal O}\bigl((x^\a)^2\bigr)$.
This allows an expansion of the metric in terms of the Riemann
tensor,
\SP{
ds^2 ~&=~ -2 du dv +\d_{ij} dy^i dy^j \\
&~~~- \bigl[R_{\a u\b u}x^\a x^\b du^2 ~+~
\tfrac{4}{3} R_{\a u\b\d}x^\a x^\b du dx^\d ~+~
\tfrac{1}{3} R_{\a\c\b\d}x^\a x^\b dx^\c dx^\d \bigr] ~+~
{\cal O}\bigl((x^\a)^3\bigr) \ .
\label{co}
}

Now return to the geodesic deviation equations \eqref{cc} and
\eqref{ce} for the transverse Jacobi fields, $z^i$ in Fermi coordinates. 
These take essentially the same form except that because of the defining
property of Fermi coordinates we can replace the covariant derivatives
along the geodesic by simple derivatives, {\it i.e.}
\EQ{
{d\over du}z^i ~=~ \W^i{}_j z^j \ , 
\label{cp}
}
\EQ{
{d^2 \over du^2} z^i ~=~ -h^i{}_j z^j \ ,
\label{cpp}
}
with $\W_{ij} = \nabla_i \hat k_j$ and $h^i{}_j = R^i{}_{uju} =
-\partial_u\W^i{}_j - \W^i{}_k \W^k{}_j$ on $\c$. 

For deviations around the preferred geodesic $\c$,
the solutions can be written in terms of the initial data at some point
$u'$ on $\c$ by introducing matrix functions $\BA(u,u')$ and $\BB(u,u')$
as follows:
\EQ{
z^i(u) ~=~ B^i{}_j(u,u') z^j(u') + A^i{}_j(u,u')\dot z^j(u') \ ,
\label{cq}
}
where $\BA(u,u')$ and $\BB(u,u')$ themselves satisfy the Jacobi 
equation
\EQ{
\ddot\BA +\Bh \BA = 0 \ , ~~~~~~~~~~~
\ddot\BB +\Bh \BB = 0 \ ,
\label{cr}
}
with boundary conditions $\BA(u',u') = 0, ~ 
\partial_u \BA(u,u')|_{u=u'} = \Bone, ~
\BB(u',u') = \Bone$, ~and
$\partial_u \BB(u,u')|_{u=u'} = 0.$ 
Clearly, the functions $\BA$ and $\BB$ carry the same information 
about the geodesic flow as the optical tensors. We can make this
explicit by combining \eqref{cc} and \eqref{cq} to give the relation:
\EQ{
\BB(u,u') + \BA(u,u') \BOmega(u') ~=~ 
\exp\left[ \int_{u'}^u du'' \big(\tfrac12\hat\theta \d_{ij} +
\hat\s_{ij}\big)\right]
\label{cs}
}
defining the optical tensors as before from $\W_{ij}$.

For our purposes, an interesting case is to define a congruence,
and therefore the optical tensors, by choosing ``geodesic spray'' 
boundary conditions such that $z^i(u') = 0$. This isolates the
function $\BA(u,u')$ and we find
\EQ{
\partial_u \log \BA(u,u') ~=~ \BOmega(u)\ .
\label{ct}
}
Taking the trace gives the identity
\EQ{
\partial_u \log\det \BA(u,u') ~=~ \hat\theta(u)\ .
\label{cu}
}

The importance of this function $\BA(u,u')$ lies in its relation to
the Van-Vleck Morette matrix and geodesic interval in a plane wave 
spacetime. Indeed, solving \eqref{cr} for $\BA$ with appropriate 
boundary conditions proves to be the quickest way to determine
the VVM matrix and the propagators in the Penrose plane wave limit.

\subsection{Penrose limit and plane wave geometry}

The Penrose limit associates with any metric $g_{\m\n}$ and null 
geodesic $\c$ a plane wave spacetime which encodes the geometry
of geodesic deviation around $\c$.
Formally, it is found by making a global Weyl transformation
followed by an asymmetric global coordinate rescaling such that 
the affine parameter along $\c$ is preserved. That is,
the Penrose limit is the metric
\EQ{
d\hat s^2 ~=~ \lim_{\l\to 0} ~\l^{-2} ds_\l^2 \ ,
\label{cv}
}
where $ds_\l^2$ is the metric \eqref{bd} with the coordinate
rescaling
\EQ{
(u,V,Y^a) ~\rta~ (u, \l^2 V, \l Y^a) \ .
\label{cw}
}
This leaves
\EQ{
d\hat s^2 ~=~ -2 du dV +C_{ab}(u) dY^a dY^b \ ,
\label{cx}
}
where $C_{ab}(u) = C_{ab}(u,0,0)$ are the transverse metric
components restricted to the geodesic $\c$ with $V=Y^a=0$.
Since, as we have just seen, these components control geodesic 
deviation and the optical tensors, it follows that the truncation 
of the original metric implied by the Penrose limit encodes 
precisely the information we need to take account of geodesic 
fluctuations in the context of QFT.

The Penrose limit \eqref{cx} is the metric for a plane wave in 
{\it Rosen coordinates}. This is a powerful result: it allows us 
with no loss of generality to consider only plane wave backgrounds 
in our analysis of photon propagation in curved spacetime.
A familiar alternative description of the plane wave metric is in 
terms of {\it Brinkmann coordinates} $(u,v,y^i)$. Around the preferred
geodesic $\c$ these are null Fermi coordinates \cite{Blau:2006ar},
which explains their importance in our analysis of vacuum polarization.
Applying the Penrose rescaling \eqref{cv},\eqref{cw} with respect to
these coordinates, the general metric \eqref{co} reduces to
\EQ{
ds^2 ~=~ -2 du dv -R_{iuju}(u) y^i y^j du^2 + \d_{ij} dy^i dy^j\ ,
\label{cy}
}
which is the well-known form for a plane wave in Brinkmann coordinates.
Note that this depends only on the curvature components $R_{iuju}$ 
which occur in the Jacobi equation. We therefore see that in both coordinate descriptions the Penrose limit captures the essential 
physics of geodesic deviation.

The relation between the Rosen coordinates $(u,V,Y^a)$ and Brinkmann
coordinates $(u,v,y^i)$ is best expressed in terms of a zweibein 
$E^i{}_a(u)$ which relates the transverse coordinates and ensures
that the transverse space is flat in Brinkmann coordinates, {\it i.e.}
\EQ{
C_{ab}(u) ~=~ E^i{}_a(u) \d_{ij} E^j{}_b(u)
\label{cz}
}
The transformation of the null coordinates is found by evaluating the 
geodesic equation in Brinkmann coordinates and using the fact that
in Rosen these are $V =$ const., $Y^a =$ const. This gives
\SP{
Y^a ~&=~ y^i E_i{}^a \ , \\
V ~&=~ v -\frac12 \W_{ij} y^i y^j \ , 
\label{cca}
}
where $E_i{}^a(u)$ is the inverse zweibein 
and we {\it define}\footnote{$\Omega_{ij}$ defined as 
$\nabla_i\hat k_j$ is clearly symmetric for a gradient field 
$\hat k^\m$. To prove symmetry directly from \eqref{ccb}, note that
the zweibein is parallel transported along the geodesic $\c$, 
{\it i.e.} $\nabla_u E^i{}_a = 0$. This condition implies
$$
\dot E^i{}_a ~=~ E^{ic} \dot E^k{}_c E_{ka} \ .
$$
Contracting appropriately with a further zweibein then shows 
either $\W_{ij} = \W_{ji}$ or $\W_{ab} = \W_{ba}$.}
\EQ{
\W^i{}_j  ~=~ E_j{}^a {d\over du} E^i{}_a  \ .
\label{ccb}
} 
Inverting,
\SP{
y^i ~&=~ E^i{}_a Y^a \ , \\
v ~&= V +\frac12 \W_{ab} Y^a Y^b \ ,
\label{ccc}
}
with $\W_{ab} = E^i{}_a \W_{ij} E^j{}_b$.\footnote{While $\W_{ab}$ 
are the only non-vanishing components of $\nabla_\m \hat k_\n$ in
Rosen coordinates, in Brinkmann we find
$$
\W_{\m\n} = \left(\begin{matrix}
y\W^3 y &0 &-y\W^2\\
0&0&0 \\
-\W^2 y &0 &\W \\
\end{matrix}\right)
$$
where on the r.h.s.~$\W$ just denotes the transverse matrix $\W_{ij}$.
Notice, using \eqref{ccg}, that this satisfies $\hat k^\m \W_{\m\n}=0$;
see footnote 3.}  Now, using the definition
\eqref{cz} to transform the Rosen form \eqref{cx} of the plane wave 
metric to Brinkmann form \eqref{cy}, we find
\EQ{
h_{ij} ~=~ R_{iuju} ~=~  -E_i{}^a {d^2 \over du^2}E_{ja}\ ,
\label{ccd}
}
consistent with the explicit result \eqref{cg} for the curvature.
Since this also shows that
$h_{ij} = -{d\over du}\W_{ij} -\W_{ik}\W^k{}_j$,
we confirm that $\W_{ij}$ defined here is the same as the previous
definition $\W_{ij} = \nabla_i \hat k_j$. This is also immediately 
apparent from \eqref{ccg} below.

\vskip0.2cm
The eikonal ansatz for the electromagnetic field is very simple in
the plane wave gravitational metric. In Rosen coordinates, the 
ansatz \eqref{ba} is realized with the tree-level eikonal phase 
$\Theta=-\w V$, which gives $\hat k^\m = (1,0, \underline 0)$,
while solving \eqref{be} for the amplitude and polarization we find
\EQ{
{\cal A} ~\propto \det \bigl( E^i{}_a \bigr)^{-\frac12} \ , 
\label{cce}
}
\EQ{
\hat\vare_{(i)a} ~=~  E_{ia} \ .
\label{ccf}
}
the latter being the solution of ~ 
$\partial_u\hat\vare_{(i)a} - \W_a{}^b \hat\vare_{(i)b} = 0$.
In Brinkmann coordinates, where 
$\Theta = \w (v + \frac12 y\W y)$,
we have the same result for the amplitude while,
\EQ{
\hat k^\m ~= \bigl(1, -y(\tfrac12 \dot\W + \W^2)y, \W_{jk}y^k\bigr) \ , 
\label{ccg}
}
\EQ{
\hat\vare_{(i)} ~=~ \bigl(-\W_{ik}y^k, 0, \d_{ij}\bigr) \ .
\label{cch}
}

\subsection{The Van Vleck-Morette matrix}

A key ingredient in the quantum field theory analysis of propagators 
and vacuum polarization in curved spacetime is the Van Vleck-Morette determinant. This is because it captures the geometry of geodesic deviation. It also has the important property of becoming singular at conjugate points along a null geodesic which, as we shall see, is the 
key to understanding the all-important analytic structure of Green
functions.

The Van Vleck-Morette matrix is defined from the geodesic interval
\EQ{
\s(x,x') ~=~ -\frac12 \int_0^1 d\t\, g_{\m\n}(x) \dot x^\m \dot x^\n \ ,
\label{cci}
}
where $x^\m(\t)$ is the null geodesic joining $x = x(0)$ and 
$x' = x(1)$, and is
\EQ{
\D_{\m\n}(x,x') ~=~ {\partial^2 \s(x,x')\over \partial x^\m ~
\partial x'^\n} \ .
\label{ccj}
}

The VVM matrix is particularly simple in the plane wave background,
primarily due to the isometry of the metric with Killing vector 
$\partial_V$. First, consider the geodesic interval in Rosen 
coordinates. We have, 
\SP{
\s(x,x') ~&=~ \frac12 (u-u') 
\int_{u'}^u du'' ~\Bigl(2\dot V - C_{ab} \dot Y^a \dot Y^b\Bigr) \\
&=~ (u-u')(V-V') -\frac12(u-u')\Bigl[Y^a C_{ab}\dot Y^b\Bigr]_{u'}^u
\ , 
\label{cck}
}
where we have used the geodesic equation 
$\ddot Y^a -2\W^a{}_b \dot Y^b = 0$ or 
${d\over du} \bigl(C_{ab} \dot Y^b\bigr) = 0$.
Solving this gives $\dot Y^a = C^{ab}(u) \xi_b$ with $\xi_b$ constant
(note that the original congruence with $Y^a = $~constant is just the
special case $\xi_b = 0$) and so
\EQ{
(Y - Y')^a ~=~ \int_{u'}^u du'' ~C^{ab}(u'')~ \xi_b \ .
\label{ccl}
}
Substituting back, we find
\EQ{
\s(x,x') ~=~ (u-u')(V-V') - \frac12 \D_{ab}(u,u') (Y-Y')^a (Y-Y')^b \ ,
\label{ccm}
}
where we define
\EQ{
\D_{ab}(u,u') ~=~ 
(u-u') ~\biggl[~ \int_{u'}^u du'' ~\BC^{-1}~\biggr]_{ab}^{-1} \ .
\label{ccn}
}
$\D_{ab}(u,u')$ are the transverse components of the Van Vleck-Morette
matrix, the determinant itself being simply
\EQ{
\det\D_{\m\n}(x,x') ~=~ - \det\D_{ab}(u,u') \ ,
\label{cco}
}
dependent only on the 2 dimensional transverse subspace.

The corresponding result in Brinkmann coordinates is similar.
Here, we have
\SP{
\s(x,x') ~&=~ \frac12 (u-u') 
\int_{u'}^u du'' ~\Bigl(2\dot v + h_{ij} y^i y^j - 
\dot y^i \d_{ij} \dot y^j \Bigr)\\
&=~ (u-u')(v-v') - 
\frac12 (u-u') \bigl[y^i\d_{ij} \dot y^j\bigr]_{u'}^u \ .
\label{ccp}
}
In the second line, we have again used the geodesic equation 
$\ddot y^i +h^i{}_j y^j = 0$. 
Since this is just the Jacobi equation \eqref{cpp}, we can use the
expansion \eqref{cq} to rewrite the geodesic interval in terms of
the $\BA$ and $\BB$ functions as follows:
\SP{
\s(x,x') ~=~ &(u-u')(v-v') - \frac12(u-u') \Bigl[
y(u)\bigl(\BA^{-1 \top}(u,u') - \BA^{-1}(u',u)\bigr) y(u') \\
&+~ y(u) \BA^{-1}(u',u) \BB(u',u) y(u)~~
- ~~y(u') \BA^{-1}(u,u') \BB(u,u') y(u') \Bigr]\ .
\label{ccq}
}
Finally, using the property $\BA^\top(u,u') = - \BA(u',u)$,
we find the transverse VVM matrix components in Brinkmann coordinates:
\EQ{
\D_{ij}(u,u') ~=~ -(u-u') A^{-1}_{ji}(u,u') \ .
\label{ccr}
}

Both these forms \eqref{ccn} and \eqref{ccr} for the transverse
VVM matrix components depend on quantities governing geodesic 
deviation. To see their equivalence, note that since both $\BA$ and
the zweibein $\BE$ satisfy the Jacobi equation, their Wronskian
is a constant matrix $\BK$, {\it i.e.}
\EQ{
\BE^\top \dot\BA - \dot\BE^\top \BA = \BK \ .
\label{ccs}
}
Integrating this equation, and imposing
the boundary conditions defining $\BA(u,u')$, we find
\EQ{
A_{ij}(u,u') ~=~ 
E_{ia}(u) ~\int_{u'}^u du'' ~C^{ab}(u'')~ E_{jb}(u') \ ,
\label{cct}
}
verifying the transpose/antisymmetry property of $\BA(u,u')$
used above.  This confirms the expected relation between
the transverse VVM matrix components in Rosen and Brinkmann, {\it i.e.}
\EQ{
\D_{ab}(u,u') ~=~ E^i{}_a(u) \D_{ij}(u,u') E^j{}_b(u') \ .
\label{ccu}
}
This allows us to find the Rosen VVM matrix directly by solving the
Jacobi equation with appropriate boundary conditions for $\BA(u,u')$,
rather than using the integral form \eqref{ccn}.
As noted at the end of section 3.1, this turns out to be the most
efficient method of evaluation in the QFT examples that follow.

\section{Refractive Index for Scalar QED}

With these geometric preliminaries complete, we now return to 
quantum field theory and the calculation of the refractive index
for photons propagating in curved spacetime. In this section, we 
review briefly the results for scalar QED derived in 
refs.\cite{Hollowood:2007kt,Hollowood:2007ku,Hollowood:2008kq};
our new results on the generalization to spinor QED are given
in sect.~5. 

\subsection{Vacuum polarization and the refractive index}

\begin{figure}[ht] 
\centerline{\includegraphics[width=2.5in]{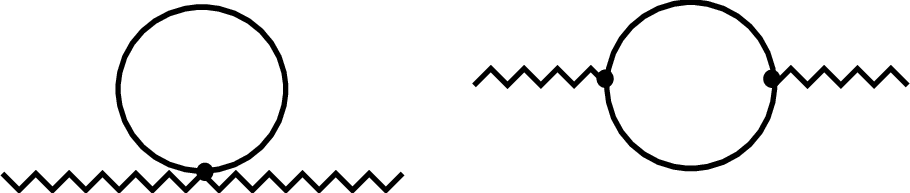}}
\caption{\footnotesize The two Feynman diagrams that contribute to the
vacuum polarization to order $\alpha$.}\label{p5}
\end{figure}

To find the refractive index at ${\cal O}(\a)$, we need to evaluate 
the r.h.s.~of \eqref{bg} with $A^\n(x')$ approximated by the
eikonal ansatz \eqref{ba} with $\Theta(x') = -\w V'$. In scalar QED, the 
one-loop vacuum polarization tensor is given by the Feynman diagrams 
shown in Fig.~\ref{p5}, viz:
\SP{
\Pi^\text{1-loop}_{\mu\nu}(x,x')&=
4\pi\a g_{\mu\nu}\delta^{(4)}(x-x')G(x,x')\\
&+8\pi\a \Big[\partial_\mu G(x,x')\partial'_\nu G(x,x')
- G(x,x')\partial_\mu\partial'_\nu G(x,x')\Big]\ .
\label{da}
}

The scalar propagator $G(x,x')$ in a general curved spacetime
can be expressed in the following form, using standard heat-kernel
or proper-time methods:
\EQ{
G(x,x')~=~
\frac{\sqrt{\det\Delta_{\mu\nu}(x,x')}}{\big(g(x)g(x')\big)^{1/4}}
\int_0^\infty\frac{dT}{(4\pi T)^2}\,ie^{-im^2T+
\tfrac1{2iT}\sigma(x,x')}~\Omega(x,x'|T)\ .
\label{db}
}
This arises from a representation of the propagator $G(x,x')$ in terms 
of a path integral over fluctuations around the classical path from
$x$ to $x'$. The VVM determinant accounts for the Gaussian terms, while
the factor $\Omega(x,x'|T)$ encodes higher corrections. It can be
expanded in the form 
$\Omega(x,x'|T) = 1 + \sum_{n=1}^\infty a_n(x,x')T^n$
where the coefficients $a_n$ are functions of the curvature and are
well-known at low order. Clearly, this translates into an expansion
in powers of $R/m^2$ in the propagator.

Now focus on the second diagram in Fig.~\ref{p5}.
Inserting the eikonal ansatz in \eqref{bg} and isolating the 
exponent term, we find
\SP{
&\int \sqrt{g(x')}\,d^4x'\,\Pi^\text{1-loop}_{\m\n}(x,x') A^\n(x')~=~\\
&~~~~~~\int \sqrt{g(x')}\,d^4x'
\int_0^\infty \frac{dT}{T^3}\int_0^1\frac{d\xi}{[\xi(1-\xi)]^2}~
e^{-im^2T + \tfrac1{2iT\xi(1-\xi)}\sigma(x,x')-i\w V'}~
\Bigl[ ~\ldots ~ \Bigr] \ ,
\label{dc}
}
where we have introduced a change of variable $T_1 = T\xi$ and
$T_2 = T(1-\xi)$ on the proper time parameters for the two propagators.
Now, in the limit $R\ll m^2$, we can evaluate the integral over $x'$ 
using a stationary phase approximation, found by extremizing this 
exponent with respect to $x'$, {\it i.e.}
\EQ{
\frac1{2T\xi(1-\xi)}\partial'_\mu\sigma(x,x')-
\omega\partial'_\mu V'=0\ .
\label{dd}
}
Since $\partial^{\prime\m}\s(x,x')$ is the tangent vector at $x'$
of the geodesic passing through $x'$ and $x$, the stationary phase 
solution corresponds to the null geodesic with tangent vector 
$\partial^{\prime \mu}V' = \hat k^\m(x')$, {\it i.e.}~the original
photon trajectory. Assuming that the background spacetime has an 
isometry $\partial_V$, we can write simply
\EQ{
\partial_{V'}\s(x,x') = -(u-u') \ ,
\label{de}
}
and so determine $u'$ from
\EQ{
u - u' ~=~ 2\w T\xi(1-\xi) \ .
\label{df}
}

This is a crucial step in our analysis. The use of the stationary
phase method at this point allows us to go beyond the expansion
in powers of $\w$ used in the all-orders derivative expansion of the effective action in refs.\cite{Shore:2002gw,Shore:2002gn}. Capturing
the non-perturbative dependence of the refractive index on $\w$ is
essential in determining its high-frequency behaviour, which in turn
is necessary for understanding how causality is realized. 

The next critical observation is that the corrections to the vacuum polarization $\Pi^\text{1-loop}_{\mu\nu}(x,x')$ at leading order in 
the weak-curvature $R/m^2$ expansion then arise from the Gaussian fluctuations around the classical null geodesic joining $x$ and $x'$.
It follows that the relevant aspects of the geometry of the 
background spacetime are simply those encoded in its Penrose limit.
Moreover, the asymmetric scaling \eqref{cw} also ensures that the stationary phase solution \eqref{df} remains invariant in this limit.
So to leading order in $R/m^2$, we only need evaluate 
$\Pi^\text{1-loop}_{\m\n}(x,x')$ in the Penrose plane wave 
corresponding to the given curved spacetime and classical photon 
trajectory.

From now on, therefore, we work exclusively in the Penrose limit.
In this case, we can adopt Rosen coordinates and use the simple 
expression \eqref{ccm} for the geodesic interval, as well as the 
solutions \eqref{cce} and \eqref{ccf} for the amplitude and 
polarization. From \eqref{bhh}, the refractive index is
\EQ{
n_{ij} ~=~ 1 + \frac{2}{\w^2} \int dx' \sqrt{g(x')}~
{{\cal A}(x')\over {\cal A}(x)}~
\hat\vare_{(i)}^\m(x) ~\Pi^\text{1-loop}_{\m\n}(x,x')~ 
\hat\vare_{(j)}^\n(x')~e^{i\w(V-V')} \ .
\label{dg}
}
For convenience, we choose the origin of coordinates so that 
$x=(u,0,0,0)$. The integral over $V'$ is trivial and, using 
\eqref{ccm}, leads to a delta function constraint
\EQ{
\int dV'\,\exp\Big[\frac{(u'-u)V'}{2iT\xi(1-\xi)}
-i\omega  V'\Big] ~=~
4\pi T\xi(1-\xi)\delta\big(u'-u+2\omega T\xi(1-\xi)\big)
\label{dh}
} 
which saturates the $u'$ integral. This automatically enforces the condition \eqref{df}, since the stationary phase solution becomes 
exact for the plane wave background.\footnote{Also note that in a plane
wave background, the form \eqref{db} of the propagator is WKB exact
with $\Omega(x,x'|T)=1$, so in this special case the leading-order
results are actually exact for all $R$, $m$ and $\w$. For a general
spacetime, our analysis applies in the limits $R\ll m^2$ and 
$\w\gg\sqrt{R}$, the latter imposed by the eikonal approximation.}

Since $\vare^\m_{(i)} =(0,0,E_i{}^a)$ is only non-vanishing in the $Y^a$ directions, the contractions in \eqref{dg} pick out just the
transverse $\Pi^\text{1-loop}_{ab}$ components. Since the only 
dependence on $Y^a$ and $Y^{\prime a}$ occurs in the exponents,
the derivatives in \eqref{da} just act on these factors
and the integrals over $Y^{\prime a}$ are simply
Gaussian.\footnote{This is a significant simplification over 
performing the whole calculation explicitly in Brinkmann coordinates, 
where the geodesic interval is \eqref{ccq} and the relation 
$$
\BB(u,u') + \BA(u,u') \BOmega(u') ~=~ 
\BE(u) \BE^{-1\top}(u')
$$
between $\BA$ and $\BB$, equivalent to \eqref{cs}, has to be used.}
These are readily evaluated in terms of the VVM determinant:
\EQ{
\int d^2Y'\,\frac{\partial}{\partial Y^{\prime a}}
e^{\tfrac i{4T\xi}Y'\cdot\Delta(u,u')\cdot Y'}\frac{\partial}{\partial
  Y^{\prime b}}
e^{\tfrac i{4T(1-\xi)}Y'\cdot\Delta(u,u')Y'}=
\frac{\pi\xi(1-\xi)}{2}\frac{\Delta_{ab}(u,u')}
{\sqrt{\det\Delta_{ab}(u,u')}}\ .
\label{di}
}

To complete the calculation, we now substitute the Rosen solutions 
for the polarizations and amplitudes, and simplify the various factors
involving the metric determinant and zweibeins. Then, the contribution
of the first Feynman diagram in Fig.~\ref{p5} is added, which removes 
the UV divergence at $T=0$. Finally, to express the refractive index 
in its most useful form, we use \eqref{ccu} to convert the VVM factors 
from Rosen back to Brinkmann. This gives the remarkably elegant result:
\SP{
\Bn^\text{scalar}(u;\omega)&=\Bone
+\frac{\alpha}{2\pi\omega^2}\int_0^{\infty}\frac{dT}{T^2}\,
ie^{-im^2T}\\ &~~~~~~~~\times
\int_0^1d\xi\,\left[\Bone
-\BDelta(u,u')\sqrt{\det\BDelta(u,u')}
\right]_{u'=u-2\omega T\xi(1-\xi)}\ 
\label{dj}
}
where $\BDelta$ is the VVM matrix in Brinkmann coordinates, {\it
  i.e.\/}~with elements $\BDelta_{ij}$.
Of course, implicitly $\Bn(u;\omega)$ is evaluated at
a point $x(u)$ on the geodesic $\gamma$.

This is our final result for the refractive index and gives the 
full frequency dependence of the phase velocity of photons travelling
through an arbitrary background spacetime. 
The key insight, that to this order the quantum effects on photon
propagation are entirely determined by the geometry of geodesic
fluctuations around the classical null trajectory and are therefore  
encoded in the plane-wave Penrose limit of the original spacetime, 
explains why the final result depends so simply on the VVM matrix only.

An immediate consistency check is to recover the low-frequency limit
of the refractive index directly from \eqref{dj} and compare with the results of section 2 obtained from the effective action.
The low-frequency behaviour is found by expanding the VVM matrices
in powers of $t = 2\w\xi(1-\xi)T$, since this is equivalent to 
expanding the refractive index itself in powers of $\w^2
R/m^4$. Substituting this expansion, 
\EQ{
\D_{ij}(u,u-t) ~=~ \d_{ij} +{1\over6}R_{uiuj}(u)t ~+~{\cal O}(t^2)\ ,
\label{dk}
}
into \eqref{dj}, we find
\EQ{
n_{ij}^\text{scalar}(u;\omega)=\delta_{ij}-\frac{\alpha}{360\pi
  m^2}\big(R_{uu}\delta_{ij}+2R_{uiuj}\big)+\frac R{m^2}{\cal O}\Big(
\frac{\omega^2R}{m^4}\Big)\ ,
\label{dl}
}
in Brinkmann coordinates, in agreement with \eqref{bo} derived from
the effective action. 

\subsection{Analyticity}

The analyticity properties of the refractive index are essential in
understanding how causality is realized for QED in curved spacetime.
We summarize here (see ref.\cite{Hollowood:2008kq} for an extensive
discussion) how the singularities in the VVM determinant 
induced by the existence of conjugate points give rise
to a novel analytic structure for the refractive index in the
complex $\omega$ plane. This makes $S$-matrix theory and dispersion
relations very different in curved spacetime from the familiar
flat spacetime axioms and theorems. 

It is convenient first to rewrite \eqref{dj} in the form
\EQ{
\Bn(u;\omega)
=\Bone-\frac{\alpha}{2\pi\omega}\int_0^1
d\xi\,\xi(1-\xi)\BcalF\Big(u;
\frac{m^2}{2\omega\xi(1-\xi)}\Big)\ ,
\label{dm}
}
where 
\EQ{
\BcalF(u;z)=
\int_0^{\infty - i\epsilon}\frac{dt}{t^2}\,ie^{-izt}\left[
\BDelta\big(u,u-t\big)\sqrt{\det\BDelta\big(u,u-t\big)}-\Bone
\right]\ ,
\label{dn}
}
Notice that we have to introduce a prescription, as indicated, for
dealing with the branch-point singularities of the integrand that  
arise whenever $x(u)$ and $x(u-t)$ are conjugate points and
$\det\BDelta(u,u-t)$ diverges. This 
prescription actually follows from a careful treatment of the VVM
determinant factor in the propagator ({\it e.g.\/}~see the book 
\cite{Kleinert}). Where we have written $\det\,\Delta_{\mu\nu}(u,u-t)$,
it should be interpreted as 
\EQ{
e^{-i\pi\nu/2}\big|\det\,\Delta_{\mu\nu}(u,u-t)\big|\ ,
}
where $\nu$ is the {\it Maslov-Morse Index\/} which counts the number
of conjugate points (or more properly the number of times an eigenvalue of
$\Delta_{\mu\nu}(x,x')$ diverges) on the geodesic joining $x(u)$ and
$x(u-t)$. Another way to interpret the phase is that it provides the
prescription for the analytic continuation of 
$\BDelta_{\mu\nu}(u,u-t)$ into the
complex $t$ plane. In particular, it requires the $t$ contour to avoid
the singularity at a conjugate point by veering into the lower-half $t$
plane as indicated above. This choice is consistent with the
requirement that the flat-space limit is smooth. 
Note that since the integral is over $\RE t >0$, it receives support
only from that part of the null geodesic to the {\it past\/} of $x(u)$,  
{\it i.e.\/}~from $\BDelta(u,u-t)$ with $\RE t >0$, as expected in a
causal theory. 

The most important observation is that 
$\BcalF(u;z)$ is by construction
guaranteed to be analytic in the lower-half $z$ plane (including the
positive real axis). Since $z$ and
$\omega$ are inversely related this means that $\Bn(u;\omega)$ is
analytic in the upper-half of the $\omega$ complex plane (including
the positive real axis). This establishes 
the first requirement for micro-causality.
The second requirement rests on the behaviour of the
large $|\omega|$ limit of the refractive index. 
Large $|\omega|$, corresponds to $z\to0$, in whose limit
\EQ{
\Bn(u;\omega)
=\Bone-\frac{\alpha}{12\pi\omega}
\BcalF(u;0)+\cdots\ ,
\label{dm2}
}
The question here is whether the $t$ integral defining $\BcalF(u;0)$ 
is convergent. If the spacetime is asymptotically flat in the far past,
then $\BDelta(u,u-t)$ asymptotes to a constant and the integral is
indeed convergent. However, this is sufficient but not necessary as we
shall see with particular examples.

Assuming that $\Bn(u;\omega)$ is analytic for large $\omega$ in the
upper-half plane then we can write the Kramers-Kronig relation in the
form \eqref{ab}, where the contour avoids any non-analyticity on the
real axis but lying just above the real axis. What we shall find is
the condition $\Bn(-\omega+i\e)=\Bn(\omega+i\e)^*$ is not satisfied in
any of the curved space cases. This should not be surprising since the
background geometry breaks Poincar\'e
invariance.

\section{Refractive Index for Spinor QED}

In our previous work \cite{Hollowood:2008kq,Hollowood:2007ku,
Hollowood:2007kt}
we have focused on scalar QED in order to investigate the novel 
physics of photon propagation in curved spacetime in a relatively 
simple context. Here, we develop the extra formalism needed to study 
the realistic case of spinor QED and derive an explicit expression 
for the full frequency dependence of the refractive index in this 
theory.

\subsection{Spinor propagator in a plane wave spacetime}

In order to define spinors in curved spacetime, we first introduce 
a local pseudo-orthonormal frame at each point in spacetime by 
means of a vierbein:
\EQ{
g_{\mu\nu}~=~e_\mu{}^A e_\nu{}^B\eta_{AB}\ ,
\label{ea}
}
where we preserve the null structure in the local frame by taking
\EQ{
\eta_{AB}=\MAT{0 & -1 & 0 & 0\\ -1 & 0 & 0 & 0\\
0&0&1&0\\ 0&0&0&1 }\ .
\label{eb}
}
For a plane wave spacetime, in Rosen coordinates, the vierbeins are explicitly
\EQ{
e^A{}_\mu~=~e_\mu{}^A~=~\MAT{{\bf 1} & 0\\ 0 & E^i{}_a}\ .
\label{ec}
}
In the local frame we denote the indices as $A=(+,-,i)$. 
Notice that the indices $i,j=1,2$ are common to the local frame and 
the Brinkmann coordinates in the transverse space. The $\gamma$ 
matrices are defined in the local frame and satisfy
\EQ{
\{\gamma^A,\gamma^B\} ~=~ -2\eta^{AB} \ .
\label{ed}
}
In particular, along the null directions the gamma matrices are
nilpotent, $\gamma^+\gamma^+ = \gamma^-\gamma^- = 0$. This is a 
crucial simplification, as we see below.

The next step is to define the spin connection. From the metric 
condition $\nabla_\m e^\n{}_B = 0$, this is:
\EQ{
\omega_{\mu\, AB} ~=~ e_{A\nu}\partial_\mu
e^\nu{}_B+e_{A\nu}\Gamma^\nu{}_{\mu\rho}e^\rho{}_B \ ,
\label{ee}
}
and has non-vanishing components
\EQ{
\omega_{a\,i+} ~=~ -\omega_{a\,+i} ~=~ -\dot E_{ia}(u)\ .
\label{ef}
}
One can verify explicitly that the connection is torsion free:
$de_A+\omega_{AB}\wedge e^B=0$, or in components,
\EQ{
\partial_{[\mu}e_{A\nu]}+\omega_{[\mu\,AB}e^B{}_{\nu]}=0\ .
\label{eg}
}
The covariant derivative on spinors is defined as
\EQ{
\nabla_\mu=\partial_\mu+\tfrac12\omega_{\mu\, AB}\sigma^{AB} \ ,
\label{eh}
}
where $\sigma^{AB}=\tfrac14[\gamma^A,\gamma^B]$. In Rosen components, 
this gives
\EQ{
\nabla_\mu=\Big(\partial_u,\partial_V,
\partial_a-\tfrac12\dot E_{ia}(u)\gamma^i\gamma^+\Big) \ ,
\label{ei}
}
so the Dirac operator is 
\EQ{
\SLASH{\nabla}=\gamma^Ae_A{}^\mu{}\nabla_\mu
=\gamma^+\partial_u+\gamma^-\partial_V-\gamma^iE_i{}^a(u)\partial_a+
\tfrac12E_i{}^a(u)\dot E_{ja}(u)\gamma^i\gamma^j\gamma^+\ .
\label{ej}
}

The spinor parallel transporter is a bi-spinor that parallel transports 
a spinor along a given path, in our case the null geodesic $\c$ joining 
$x$ and $x'$. It satisfies the two conditions
\EQ{
\partial^\mu\sigma(x,x')\nabla_\mu{\mathbb U}(x,x')=0\ ,\qquad {\mathbb
  U}(x,x)={\mathbb I}\ ,
\label{ek}
}
and has an explicit representation in terms of the spin connection by 
the path ordered expression
\EQ{
{\mathbb U}(x,x')={\cal P}\exp-\frac12\int_{u'}^u \omega_{\mu
  AB}\sigma^{AB}\,dx^\mu\ ,
\label{el}
}
where the integral is taken along $\c$. 
The key observation is that since the spin connection involves the
nilpotent $\gamma^+$, we can expand the exponential and only terms 
up to linear order in the spin connection contribute (in particular 
the issue of path ordering is moot):
\EQ{
{\mathbb U}(x,x')
={\mathbb I}+\tfrac12\gamma^i\gamma^+
\int_{Y^{\prime a}(u')}^{Y^a(u)}
\dot E_{ia}(u)\,dY^a\ .
\label{em}
}
To evaluate this, write 
\EQ{
\int_{Y^{\prime a}}^{Y^a}
\dot E_{ia}(u)\,dY^a=\int_{u'}^u
du\,\dot E_{ia}(u)\,\dot Y^a(u)\ .
\label{en}
}
and, as in section 3.3, use the geodesic equation 
$\dot Y^a = C^{ab}\xi_b$.
Since $\BC=\BE^\top\BE$ and $\BOmega=\dot\BE\BE^{-1}$ are both 
symmetric, we can write
\SP{
&\int du\,\dot\BE \BC ~=~ \int du\,\dot\BE\BE^{-1}
\big(\BE^{-1}\big)^\top
~=~ \int du\,\big(\BE^{-1}\big)^\top\dot\BE^\top
\big(\BE^{-1}\big)^\top\\ 
&=-\int du\,\big(\dot\BE^{-1}\big)^\top ~=~ 
-\big(\BE^{-1}\big)^\top\ .
\label{eo}
}
Hence 
\SP{
\int_{Y^{\prime a}}^{Y^a}
\dot E_{ia}(u)\,dY^a ~&=~ -\big(E_i{}^a(u)-E_i{}^a(u')\big)\xi_a\\ 
&=~ -\big(E_i{}^a(u)-E_i{}^a(u')\big)\frac{\Delta_{ab}(u,u')}{u-u'}
(Y-Y')^b \ ,
\label{ep}
}
using \eqref{ccl} and \eqref{ccn}. This gives the following explicit 
form for the spinor parallel transporter in terms of the VVM matrix:
\EQ{
{\mathbb U}(x,x')
={\mathbb I}-\tfrac12\gamma^i\gamma^+
\big(E_i{}^a(u)-E_i{}^a(u')\big)\frac{\Delta_{ab}(u,u')}{u-u'}
(Y-Y')^b\ .
\label{eq}
}

Finally, we need the Feynman propagator ${\mathbb S}(x,x')$ for the 
spinor electron. This can be written in terms of the propagator 
$G(x,x')$ for a massive scalar field and the spinor parallel 
transporter ${\mathbb U}(x,x')$ as follows:
\EQ{
{\mathbb S}(x,x')=\big(\SLASH{\nabla}-m\big)G(x,x'){\mathbb U}(x,x')\ .
\label{er}
}

\subsection{Vacuum polarization in spinor QED}

With this form for the spinor propagator in a plane wave spacetime, 
we can now calculate the vacuum polarization and refractive index 
following the method of section 4. The vacuum polarization is given 
by the usual one-loop Feynman diagram (the second in Fig.~\ref{p5}), 
{\it viz.}
\EQ{
\Pi^\text{1-loop}_{\mu\nu}(x,x') ~=~ e^2\text{Tr}
\big[\gamma_\mu{\mathbb S}(x,x')
\gamma_\nu{\mathbb S}(x',x)\big]\ .
\label{es}
}

Inserting the vacuum polarization for spinor QED into \eqref{bhh}
for the one-loop contribution to the refractive index, we find, 
analogously to \eqref{dg} in the scalar case,
\EQ{
n_{ij}(x;\omega)=\delta_{ij}+\frac{2e^2}{\omega^2}\int d^4x'\, 
\sqrt{g(x')}\,\frac{{\cal A}(x')}{{\cal A}(x)}~
\text{Tr}\big[\gamma_i{\mathbb S}(x,x')\gamma_j{\mathbb S}(x',x)\big]~
e^{i\omega(V - V')}\ .
\label{et}
}
Again, the calculation is best performed in Rosen coordinates 
although the end result is most naturally expressed in terms of 
tensors with Brinkmann indices. We again take $x=(u,0,0,0)$
for convenience.

Now notice that the terms which are linear in $m$ involve an odd number 
of gamma matrices and so do not contribute to the trace. There are two remaining contributions to the trace of the form
\SP{
&\text{Tr}\Big[\gamma_i\, \SLASH{\nabla} \left( G(x,x')\mathbb{U}(x,x')
\right)
\gamma_j\,\SLASH{\nabla}' \left( G(x',x)\mathbb{U}(x',x) \right)\Big]\\
&+m^2\text{Tr}\Big[\gamma_iG(x,x')\mathbb{U}(x,x')\gamma_j
G(x',x)\mathbb{U}(x',x)\Big]\ .
\label{eu}
}
The explicit expression for the spinor parallel transporter is
given in \eqref{eq}. The $Y^a$ integrals are Gaussian and easily
evaluated while the $V'$ integral is trivial and simply 
produces a delta function constraint
which enforces the same condition \eqref{df} as in the scalar case. 

Noting that 
\EQ{
\SLASH{\nabla}\mathbb{U}(x,x')\SLASH{\nabla}'\mathbb{U}(x',x)=0\ ,
\label{ew}
}
due to the fact that $\gamma^+\gamma^+=0$, there are three types of
contribution in \eqref{eu} which, suppressing the calculational
details which are not in themselves enlightening, yield:
\SP{
&\int d^4x'\,\sqrt{g(x')}\,e^{-i\omega V'}\,\frac{{\cal A}(x')}{{\cal A}(x)}
\,
\text{Tr}\Big[\gamma_iG(x,x')\mathbb{U}(x,x')\gamma_j
G(x',x)\mathbb{U}(x',x)\Big]\\ &=64\pi
T^2\xi^2(1-\xi)^2\sqrt{\det\BDelta(u,u')}\ ,
\label{ex}
}
along with
\SP{
&\int d^4x'\,\sqrt{g(x')}\,e^{-i\omega V'}\,\frac{{\cal A}(x')}{{\cal A}(x)}
\,\text{Tr}\Big[\gamma_i\, \SLASH{\nabla}G(x,x')\mathbb{U}(x,x')
\gamma_j\,\SLASH{\nabla}' G(x',x)\mathbb{U}(x',x)\Big]
\\ &=16i\pi T\xi(1-\xi)\sqrt{\det\BDelta(u,u')}
\Big\{\Big(-8\xi(1-\xi)\\ &-(1-2\xi(1-\xi))(u-u')
\text{Tr}\,\big(\partial_{u'}\BDelta(u,u')\BDelta^{-1}(u,u')\big)\Big)
\delta_{ij}+4\xi(1-\xi)\Delta_{ij}(u,u')\Big\}
\label{ey}
}
and
\SP{
&\int d^4x'\,\sqrt{g(x')}\,e^{-i\omega V'}\,\frac{{\cal A}(x')}{{\cal A}(x)}
\,
\text{Tr}\Big[\gamma_i\, \SLASH{\nabla}G(x,x')\mathbb{U}(x,x')
\gamma_j\, G(x',x)\SLASH{\nabla}'\mathbb{U}(x',x)\\ &~~~~~~+
\gamma_i\, G(x,x')\SLASH{\nabla}\mathbb{U}(x,x')
\gamma_j\,\SLASH{\nabla}' G(x',x)\mathbb{U}(x',x)\Big]
\\ &=16i\pi T\xi(1-\xi)\delta_{ij}\sqrt{\det\BDelta(u,u')}\\
&~~~~~~\times\text{Tr}\,\Big(
\Bone-\BDelta(u,u')+(u-u')\partial_{u'}\BDelta(u,u')\BDelta^{-1}(u,u')
\Big)\ .
\label{ez}
}
In all these expression $u'$ is constrained via \eqref{df}.

Putting everything together, the final result for the refractive 
index is 
\SP{
\Bn^\text{spinor}(u;\omega)&=\Bone
-\frac{\alpha}{2\pi\omega^2}\int_0^{\infty}\frac{dT}{T^2}\,
ie^{-im^2T}\\ &\times
\int_0^1d\xi\,\sqrt{\det\BDelta(u,u')}\Big[\Big(4iTm^2+\frac1{\xi(1-\xi)}
\text{Tr}\,(\BDelta(u,u')-\Bone)\\ &-2(u-u')
\text{Tr}\,(\partial_{u'}\BDelta(u,u')\BDelta^{-1}(u,u'))
+8\Big)\Bone
-4\BDelta(u,u')\Big]_{u'=u-2\omega T\xi(1-\xi)}\\
\label{eaa}
}
or equivalently,
\SP{
\Bn^\text{spinor}(u;\omega)&=\Bone
-\frac{\alpha}{2\pi\omega^2}\int_0^{\infty}\frac{dT}{T^2}\,
ie^{-im^2T}\\ &\times
\int_0^1d\xi\,\sqrt{\det\BDelta(u,u')}\Big[\Big(\frac1{\xi(1-\xi)}
\text{Tr}\,(\BDelta(u,u')-\Bone)\\ &-4(u-u')
\text{Tr}\,(\partial_{u'}\BDelta(u,u')\BDelta^{-1}(u,u'))
+4\Big)\Bone
-4\BDelta(u,u')\Big]_{u'=u-2\omega T\xi(1-\xi)}
\label{eab}
}
In the second expression, we have taken the first term and 
integrated by parts in $T$ and ignored the singular boundary term
which is curvature independent. The result is then manifestly
UV finite, {\it i.e.}~there is no singularity of the 
integrand at small $T$.

Finally, changing variables to $t=2\w\xi(1-\xi)T$ and writing the
refractive index in the form of section 4.2, we have 
\EQ{
\Bn^\text{spinor}(u;\omega)
=\Bone-\frac{\alpha}{2\pi\omega}\int_0^1
d\xi\,\xi(1-\xi)\BcalF^\text{spinor}\Big(u;
\frac{m^2}{2\omega\xi(1-\xi)}\Big)\ ,
\label{eac}
}
where 
\SP{ 
\BcalF^\text{spinor}(u;z)&=\int_0^{\infty-i\epsilon}\frac{dt}{t^2}\,ie^{-izt}
\sqrt{\det\,\BDelta(u,u-t)}\Big[\Big(\frac1{\xi(1-\xi)}
\text{Tr}\,(\BDelta(u,u-t)-\Bone)\\ &+4t\,
\text{Tr}\,(\partial_{t}\BDelta(u,u-t)\BDelta^{-1}(u,u-t))
+4\Big)\Bone
-4\BDelta(u,u-t)\Big] \ .
\label{eae}
}

The generic analyticity properties are exactly as described for scalar 
QED. Also as in the scalar case, we can verify the leading low-frequency
term in the expansion of $\Bn(u;\w)$.  Substituting the expansion
\eqref{dm} of the VVM matrix into \eqref{eae}, we find
\EQ{
n^\text{spinor}_{ij}(u;\omega)~=~\delta_{ij}-\frac{\alpha}{180\pi
  m^2}\big(13\delta_{ij}R_{uu}-4R_{uiuj}\big)
~~+~\frac {R}{m^2}{\cal O}\Big(\frac{\omega^2R}{m^4}\Big)\ ,
\label{eaf}
}
in agreement with the original Drummond-Hathrell formula \eqref{bp}.

\section{Homogeneous plane waves}

The first class of examples we shall study in detail are the
homogeneous plane waves. These are already sufficient to display many 
of the interesting consequences of curved spacetime geometry on the
analytic and causal structure of the refractive index. However, as we 
shall see, they are not just simplified toy models but actually arise as the Penrose limits of physically interesting spacetimes.

Homogeneous plane waves are characterized by an enhanced degree of 
symmetry. Any plane wave admits a Heisenberg algebra as its spacetime isometry (though this is not necessarily a symmetry of the original 
metric for which the plane wave is a Penrose limit.)
However, there are two simple cases (see \cite{Blau2} for a more 
complete classification) where this isometry group is extended -- the
symmetric plane waves, with profile function $h_{ij}$ constant, and the singular homogeneous plane waves, where $h_{ij} = c_{ij}/u^2$.
The first case is evidently invariant under $u$-translations; the
second under scale transformations $u \rta \l u,~ v \rta \l^{-1}v$.

An important insight in the study of Penrose limits is the idea of
hereditary properties, {\it i.e.}~some feature of the original metric 
which is preserved in the Penrose limit \cite{Blau2}. Amongst such hereditary properties are~  (i)~ Ricci flat (which implies 
$\text{tr}\,h_{ij}$ =0),~(ii)~conformally, or Weyl, flat ($h_{ij} = h(u)\d_{ij}$, so the Penrose limit is cylindrically symmetric),~(iii)~locally symmetric, 
{\it i.e.}~the covariant derivative of the Riemann tensor vanishes.
Another useful property is that an Einstein metric has a Penrose limit
which is Ricci flat.

Consider first maximally symmetric spacetimes, including de Sitter
(with isometry SO(1,4)) and anti de Sitter (SO(2,3)). The symmetry
implies that the Riemann tensor has the form $R_{\m\n\l\r} = 
{1\over6}\bigl(g_{\m\l}g_{\n\r} - g_{\m\r}g_{\n\l}\bigr)$,
which implies this is both an Einstein space and conformally flat.
Since this requires the Penrose limit (which is unique because the
maximal symmetry implies all geodesics are equivalent) to be both
Weyl and Ricci flat, $h_{ij}$ must be proportional to $\d_{ij}$ and
traceless, and therefore vanish. That is, the Penrose limit of a
maximally symmetric spacetime is flat.\footnote{We can derive these
properties very directly in the Newman-Penrose formalism. 
(See refs.\cite{Shore:1995fz,Shore:2003jx,Shore:2003zc,
Shore:2007um} for an extensive discussion of the NP formalism applied 
to superluminal photon propagation in curved spacetime.) As explained
further in section 7, the NP tetrad provides a realization of the
local null Fermi coordinate basis, with the Penrose limit profile 
function being determined by $R_{uiuj}$, that is by
$R_{\m\n\l\r}\ell^\m m^\n \ell^\l m^\r$ and
$R_{\m\n\l\r}\ell^\m m^\n \ell^\l \bar m^\r$,
or equivalently $C_{\ell m \ell m} = -\Psi_0$ and 
$R_{\ell \ell} = -2\Phi_{00}$. For a maximally symmetric space with 
$R_{\m\n\l\r} = {1\over6}\bigl(g_{\m\l}g_{\n\r} - g_{\m\r}g_{\n\l}\bigr)$,
$R_{\ell m \ell m}$ is proportional to $\ell^2 m^2 - \ell.m \ell.m$ 
(similarly for $R_{\ell m \ell \bar m}$), which vanishes
by the defining properties of the NP tetrad. Both $\Phi_{00}$ and
$\Psi_0$ vanish and the Penrose limit is flat. This formalism also makes
it clear why the Penrose limit of an Einstein space is Ricci flat:
if $R_{\m\n} = \L g_{\m\n}$ in the original space, then 
$\Phi_{00} = -{1\over2}R_{uu} = 
-{1\over2}R_{\m\n}\ell^\m \ell^\n \sim \ell^2 = 0$.}
Physically, this means that the refractive index receives no quantum corrections for photon propagation in de Sitter or anti de Sitter space.

Next, consider locally symmetric spacetimes, {\it i.e.}~with
$\nabla_\s R_{\m\n\l\r} = 0$. Since this is an inherited property, the
Penrose limit has a profile function $h_{ij}$ which is independent 
of $u$. This defines a {\it symmetric plane wave}. With no loss of
generality, we can diagonalize $h_{ij}$ and write the metric in
Brinkmann coordinates as 
\EQ{
ds^2 ~=~ - 2du dv - h_{ij}y^i y^j du^2 + dy^i dy^i \ ,
\label{fa}
}
where $h_{ij} = \text{diag}(\s_1^2,\s_2^2)$ with the $\s_i$ constant.
The corresponding curvatures are then $R_{uu} = \s_1^2 + \s_2^2$
and $R_{uiui} = \s_i^2$. There are two generic cases to consider:
(i)~ $\s_i$ both real (which includes the conformally flat case
$\s_1 = \s_2$) and ~(ii)~$\s_1$ real and $\s_2 = i |\s_2|$ pure 
imaginary (including the Ricci flat case $\s_1 = |\s_2|$). 
The third case, with both $\s_1$ and $\s_2$ imaginary is disallowed 
by the null energy condition, which requires $R_{uu} \ge 0$.

Symmetric plane waves arise as the Penrose limit for critical photon 
orbits in Schwarzschild spacetime, as shown in section 7, and also
in higher dimensions as the Penrose limits corresponding to a 
class of null geodesics in the AdS$_5\times S^p$ spaces
of interest in the AdS/CFT correspondence. The Penrose limit for 
geodesics lying entirely within the AdS or $S^p$ subspaces are flat,
since these subspaces are maximally symmetric. However, a null geodesic
in AdS with a non-vanishing angular momentum in the $S^p$ subspace
has a non-trivial Penrose limit. Since the original space is locally
symmetric, and this is a hereditary property, the Penrose limit for these
geodesics is a symmetric plane wave and in fact \cite{Blau2} can be 
shown to have $h_{ij} \sim {\rm diag}\bigl(\s_1^2, \ldots,\s_1^2,
\s_2^2,\ldots \s_2^2\bigr)$, with four factors of $\s_1^2$ and
$p-1$ of $\s_2^2$. (Here, $r_1 = \s_1^{-1}$ and $r_2 = \s_2^{-1}$ are
the curvature radii of the AdS and $S^p$ subspaces respectively.) 

The second class of homogeneous plane waves considered here are the
{\it singular homogeneous plane waves}, with Brinkmann metric 
\EQ{
ds^2 ~=~ -2 du dv - c_{ij}y^i y^j {du^2 \over u^2} + 
\d_{ij}dy^i dy^j\ ,
\label{fb}
}
where, again, with no loss of generality we can take $c_{ij} 
= \left(\begin{matrix} c_1 & 0 \\ 0 & c_2 \end{matrix}\right)$
to be diagonal. Here, we find that the singularity structure of
the VVM matrix, and consequently the analytic properties of the
refractive index, turn out to be dependent on the actual numerical
values of the constants $c_i$. These metrics arise as the Penrose 
limits associated with spacetime singularities, with the values 
of $c_i$ being related to the Szekeres-Iyers \cite{Szekeres1,Szekeres2}
classification of power-law singularities \cite{Blau2}. In particular,
the singular homogeneous plane waves arise as Penrose limits both
of cosmological FRW spacetimes and of the near-singularity limits of
black hole spacetimes.

\subsection{Symmetric plane waves}

Now consider the symmetric plane waves, with Brinkmann metric \eqref{fa}.
The Jacobi equation \eqref{cpp} for the connecting vector $y^i$ is 
simply
\EQ{
{d^2\over du^2} y^i+\s_i^2 y^i=0 \ ,
\label{fc}
}
with general solution $y^i(u) = a \cos(\s_i u + b)$. The VVM matrix is 
then found from the matrix $\BA(u,u')$ which solves the Jacobi equation
\eqref{cr} subject to the appropriate boundary conditions. This selects
the solution
\EQ{
A_{ij}(u,u')=\delta_{ij}\frac{\sin\sigma_i(u-u')}{\sigma_i} \ .
\label{fd}
}
so we find
\EQ{
\Delta_{ij}(u,u')=\delta_{ij}\frac{\sigma_i(u-u')}
{\sin\sigma_i(u-u')}\ . 
\label{fe}
}

Many of the key features of the analytic structure of the refractive
index are already evident in the simple case of a conformally flat
plane wave, $\s_1 = \s_2 = \s$. Recall that the refractive index is 
given in terms of the auxiliary function $\BcalF(u;z)$ by \eqref{dm}.
For scalar QED, inserting the formula \eqref{fe} for the VVM matrix
into the definition \eqref{dn}, we find 
$\BcalF(u;z)= {\cal F}(z){\bf 1}$ with
\EQ{
{\cal F}^\text{scalar}(z) ~=~ \int_0^{\infty - i\e} {dt\over t^2} i e^{-izt}
\biggl[\Bigl({\s t\over \sin \s t}\Bigr)^2-1\biggr] \ ,
\label{ff}
}
while for spinor QED, \eqref{eae} gives
\EQ{
{\cal F}^\text{spinor}(z) ~=~2 
\int_0^{\infty-i\epsilon}\frac{dt}{t^2}\,ie^{-izt}\frac{\sigma
  t}{\sin\sigma t}\Big[2\Big(3-\sigma t
\frac{2\cos\sigma t+1}{\sin\sigma t}\Big)
+\frac1{\xi(1-\xi)}\Big(\frac{\sigma t}{\sin\sigma t}-1\Big)\Big]\ .
\label{fg}
}
The integrands here have a series of poles (a special property of
the conformally flat case; in general these will be branch points)
on the positive real axis at $t = n\pi/s, ~ n = 1,2,\ldots$ corresponding
to the singularities in the VVM matrix related to conjugate points
on the classical null
geodesic\cite{Hollowood:2008kq,Hollowood:2007ku,Hollowood:2007kt}. 

The refractive index itself is then evaluated by performing the
integral over $t$, with the appropriate contour rotation described in
section 4. In this case, the refractive index is real,  
$\IM n(\w) = 0$, since the contour can be rotated $t\to-it$ to lie
along the negative imaginary axis in which case the integrand is
manifestly real. In both cases the $t$ integral can be performed
explicitly to give, for scalar QED (a result from
\cite{Hollowood:2008kq})\footnote{Here, $\psi(z)=\partial_z\Gamma(z)$ is
  the di-gamma function.} 
\EQ{
{\cal F}^\text{scalar}(z) = z\log(\tfrac{z}{2\sigma})- z\,\psi(1+\tfrac
z{2\sigma})+\sigma \ ,
\label{fh}
}
while for spinor QED
\SP{
{\cal F}^\text{spinor}(z)& = 8\big(-\sigma-3z+3\sigma\log(2\pi)-
6\sigma\log\Gamma(\tfrac12+\tfrac z{2\sigma})\\ &~~~~~~~~+
z\psi(1+\tfrac z{2\sigma})+2z\psi(\tfrac12+\tfrac z{2\sigma})\big)\\
&+\frac4{\xi(1-\xi)}\big(\sigma+z-\sigma\log(2\pi)+
2\sigma\log\Gamma(\tfrac12+\tfrac z{2\sigma})-z\psi(1+
\tfrac z{2\sigma})\big)\ .
}
In both cases ${\cal F}(z)$ is analytic in the lower half plane
(including the positive real axis); in fact the only points of
non-analyticity are at $z=0,-\sigma,-2\sigma,\ldots$, 
being either branch points
or poles, or combinations thereof.

\begin{figure}[ht]
\centerline{\includegraphics[width=2.5in]{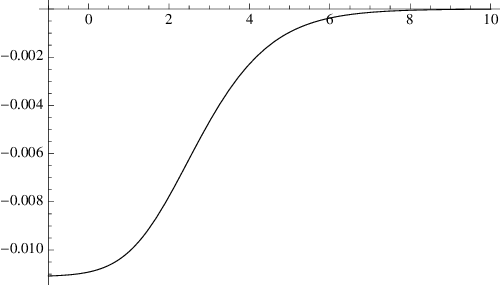}
\hspace{0.2cm}\includegraphics[width=2.5in]{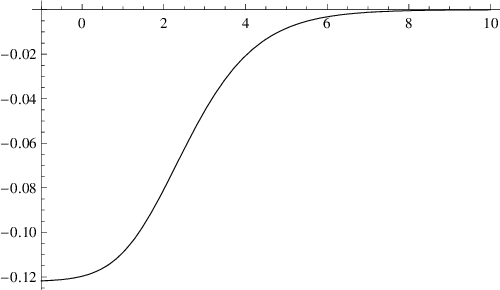}}
\caption{\footnotesize The refractive index $n(\omega)-1$ of a 
  conformally flat symmetric plane wave, in units of 
  $\alpha\sigma^2/(2\pi m^2)$, plotted as a function of 
  $\log\omega\sigma/m^2$ for scalar QED (left) and spinor QED (right).}
\label{f1}
\end{figure}
The remaining $\xi$ integrals for both scalar and spinor QED can be performed
numerically and the results for the refractive index are plotted in 
Fig.~\ref{f1}.
This shows clearly how the occurrence of a superluminal low-frequency
phase velocity is compatible with causality, 
$n(\infty) \rta 1$. Also note that since $\IM n(\w) = 0$, the 
Kramers-Kronig dispersion relation in its conventional form \eqref{aa} cannot 
hold since it is clear that $n(\w +i\e) \neq n(-\w + i\e)^*$ for
$\omega\in\R^+$. However, the more general dispersion relation
\eqref{ab} does hold as 
demonstrated in full detail in ref.\cite{Hollowood:2008kq}.

\vskip0.2cm
Another interesting example is the Ricci flat symmetric plane wave,
$\sigma_1=\sigma$ and $\sigma_2=i\sigma$. In this case, the VVM
matrix has components
\EQ{
\D_{11}(u,u') = {\s(u-u')\over\sin\s(u-u')},~~~~~~~~~~
\D_{22}(u,u') = {\s(u-u')\over\sinh\s(u-u')}\ ,
\label{fi}
}
with corresponding expressions for $\BcalF(z)$ for scalar and spinor QED.
The $t$ integrands therefore have series of branch points on both the real 
and imaginary axes. 
As a result, when the rotation of the $t$ contour, $t\to-it$ is made
so that it lies along the 
negative imaginary axis, we encounter the new branch points, 
which contribute a non-vanishing imaginary part to the refractive index. 
(See ref.\cite{Hollowood:2008kq} for full details and discussion of the
analytic structure of $\BcalF(z)$.) 

\begin{figure}[ht]
\centerline{\includegraphics[width=2.5in]{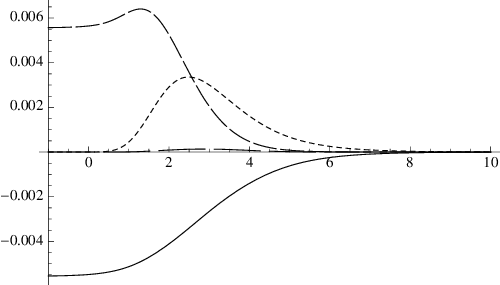}
\hspace{0.2cm}\includegraphics[width=2.5in]{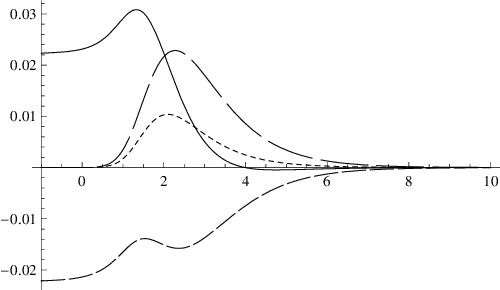}}
\caption{\footnotesize The refractive index $n(\omega)-1$ of a 
  Ricci flat symmetric plane wave, in units of 
  $\alpha\sigma^2/(2\pi m^2)$, plotted as a function of 
  $\log\omega\sigma^2/(2\pi m^2)$:
  continuous (real part, polarization $i=1$); big dashes (imaginary
  part, $i=1$); small dashes (real part, polarization $i=2$) 
  and dots (imaginary part, $i=2$), for scalar QED (left) and 
  spinor QED (right).}
\label{f3}
\end{figure}

The numerical results for the refractive index are plotted in 
Fig.~\ref{f3}. Consistent with the low-frequency results 
\eqref{bo}, \eqref{bp}, the two polarizations give opposite sign
corrections to the refractive index (gravitational birefringence), 
in contrast to the conformally flat case. The imaginary parts are
both positive for scalar and spinor QED, as expected from the flat spacetime
identification of imaginary parts of forward scattering amplitudes
with cross sections (optical theorem). In the following, however, we shall
meet situations with negative imaginary parts of the
refractive index which remain to be fully understood.

\subsection{Singular homogeneous plane waves}
\label{shpw}

We now come to the singular homogeneous plane waves, with Brinkmann 
metric \eqref{fb}. Here, the Jacobi equation is
\EQ{
{d^2\over du^2}y^i+{c_i\over u^2} y^i=0\ ,\qquad
c_i=\tfrac14(1-\alpha_i^{2})\ .
\label{ffa}
}
and it is convenient to define $\alpha_i=\sqrt{1-4c_i}$ so that the
general solution is $y^i = C u^{(1+\alpha_i)/2} + D u^{(1-\alpha_i)/2}$. 
To select the (diagonal) 
matrix $A_{ij}$, we impose the familiar boundary conditions 
$\BA(u',u') = 0$, ${d\over du}\BA(u,u')|_{u=u'} = \Bone$ 
which gives
\EQ{
A_{ii}(u,u') ~=~ \alpha_i^{-1}(uu')^{(1-\alpha_i)/2}
\bigl( u^{\alpha_i} -  u^{\prime\, \alpha_i}\bigr), ~~~~~~~ i = 1,2.
\label{ffb}
}
The VVM matrix $\BDelta(u,u')$ therefore has components 
\EQ{
\D_{ij} ~=~ \delta_{ij}\alpha_i 
(uu')^{( \alpha_i-1)/2}
\frac{u-u'}{u^{\alpha_i} -  
u^{\prime\,\alpha_i}}, ~~~~~~~ i = 1,2.
\label{ffc}
}
Notice in this case, there are no conjugate points even when
$\alpha_i$ is real. However, when $c_i>\tfrac14$, $\alpha_i$ purely
imaginary there are an infinite sequence of conjugate points when
\EQ{
u=u'\exp\left(\frac{2\pi n}{|\alpha_i|}\right)\ ,\qquad
n=1,2,\ldots\ .
}
 
Substituting into \eqref{dj} for the refractive index in scalar QED
determines the function 
${\BcalF}(z) = {\cal F}_{(i)}(z) \d_{ij}$ as 
\EQ{
{\cal F}_{(1)}(z) ~=~ \int_0^{\infty-i\e}{dt\over t^2}i e^{-izt}
\Bigl[\D_{11}^{{3/2}}(u,u-t) \D_{22}^{{1/2}}(u,u-t)-1
\Bigr] \ ,
\label{ffd}
}
for the polarization corresponding to $i=1$ (similarly for $i=2$), 
with the related, more complicated form \eqref{eac} for spinor QED. 
However, the expression above only makes sense when the singularity at
$u=0$ lies in the future of the point $u$. In other words $u<0$. This
is what happens in the case of black hole as we describe in the next
section. On the other hand, for a cosmological spacetime the
singularity occurs in the past of the point $u$. In this case the
expression above does not make sense unless the $t$ integral is
cut-off at some point $u_0>0$. This will discussed in more detail in
Section \ref{cosmo}.

In the former case, we can easily calculate the first few terms in the
frequency expansion of the refractive index $n_{ij}=n_i\delta_{ij}$:
\EQ{
n_1(u;\omega)=1-\frac{3c_1+c_2}{360\pi
  m^2u^2}+i\frac{\alpha\omega(3c_1+c_2)}{840\pi m^4u^3}+\cdots\ .
}
with a similar expression for the other polarization.

\section{Black Holes and their Singularities}

After studying these especially simple examples, we now turn to 
spacetimes arising as physically important solutions of the Einstein 
field equations and begin with black holes. This immediately raises
the question of the effect of horizons and singularities on the
causality and analyticity properties of photon propagation.
The first step is to determine the Penrose limit relevant for 
various classical trajectories. It turns out that we can do this in 
considerable generality for black hole spacetimes. Remarkably, the
near-singularity limits turn out to be just the singular homogeneous
plane waves discussed above.

\subsection{Penrose limit for Petrov type D metrics}

The relation between the Penrose limit and the geometry of geodesic
deviation explored in section 3 provides a very efficient way of
determining the profile function $h_{ij}$ of the equivalent plane wave.
The starting point is the identification of the parallel transported
frame $E_A$ introduced in \eqref{cl} in the determination of null
Fermi coordinates. The component $E_u{}^\mu = \hat k^\mu$ is simply
the tangent vector along the null geodesic $\c$; in fact, we then 
only need the explicit form for the transverse spatial components 
$E_i,~ i = 1,2$ (related to the subspace $\hat{\cal V}$ of footnote 3).
Assume now that $\c$ is just one integral curve of a known vector 
field $\hat k^\m(x)$, or equivalently assume a particular embedding
of $\c$ into some null congruence. Having identified the transverse subspace, we can immediately compute the now familiar matrix 
$\W_{ij} = \nabla_i \hat k_j$, first introduced through the Lie 
derivative \eqref{cb}, which defines the optical tensors controlling
geodesic flow. The profile function of the Penrose plane wave metric,
in Brinkmann coordinates, is then given by the identity $h_{ij} = 
\dot\W_{ij} + \W_{ik} \W^k{}_j$.

In section 7.2, we carry through this construction explicitly for 
general planar orbits in the Schwarzschild metric. First, we present 
an alternative derivation which exploits the Newman-Penrose description
of the whole class of Petrov type D spacetimes \cite{Chandra}, which
includes the Schwarzschild, Reissner-Nordstr\"om and Kerr black hole
solutions. This makes very clear how the plane wave profile function
depends on the conserved integrals of motion characterizing the
classical trajectory $\c$.

The Newman-Penrose (NP) formalism uses a null tetrad 
$E_A{}^\m = (L^\m, N^\m, M^\m, \bar M^\m)$, where the null vectors are
defined so that $L\cdot N = -1,~M\cdot\bar M= 1$, with all other scalar
products vanishing. The ten independent components of the Weyl tensor
are 
represented by the five complex scalars:
\SP{
&\hat\Psi_0 ~=~ - C_{\m\n\l\r} L^\m M^\n L^\l M^\r \\
&\hat\Psi_1 ~=~ - C_{\m\n\l\r} L^\m N^\n L^\l M^\r \\
&\hat\Psi_2 ~=~ - C_{\m\n\l\r} L^\m M^\n \bar M^\l N^\r \\
&\hat\Psi_3 ~=~ - C_{\m\n\l\r} L^\m N^\n \bar M^\l N^\r \\
&\hat\Psi_4 ~=~ - C_{\m\n\l\r} N^\m \bar M^\n N^\l \bar M^\r  \ .
\label{ga}
}
and we also encounter the scalar $\hat\Phi_{00}= -{1\over2}R_{LL}$
from the set describing the Ricci tensor.
The special feature of Petrov type D spacetimes is that we can choose
a special NP basis $(\ell^\m, n^\m, m^\m,\bar m^\u)$ where $\ell^\m$
is tangent to the {\it principal null geodesics} such that the only
non-vanishing component of the Weyl tensor is $\Psi_2$. (We reserve the
un-hatted notation for the components with respect to the principal
null geodesic basis.) 

To apply this formalism here, we note that along a given null geodesic
$\c$ with tangent vector $L^\m \equiv \hat k^\m$, we can identify the
NP frame with that describing the null Fermi coordinates. (Of course,
in the NP basis the two transverse spatial components are complex linear combinations of those in the Fermi basis.)  Determining the Penrose limit
then simply involves finding the components $R_{LMLM}$ and 
$R_{LML\bar M}$ of the Riemann tensor, or equivalently $R_{LL}$ and 
$C_{LMLM}$ (see below), since this gives the profile function of the
Penrose plane wave. 
 
It follows immediately that for photons following a principal null
geodesic, the Penrose limit is flat, since by definition $\Psi_0 = 0$.
This shows immediately that at ${\cal O}(R/m^2)$, for all frequencies,
there is no quantum correction to the refractive index for photons 
following radial geodesics in Schwarzschild spacetime, with a similar result for the corresponding principal null directions in the Kerr
or Reissner-Nordstr\"om metrics. 

To extend this to an arbitrary classical trajectory, we first expand the
corresponding tangent vector $L^\m$ in terms of the NP basis adapted to
the principal null geodesics, {\it i.e.}
\EQ{
L^\m ~=~ \a \ell^\m + \b n^\m + \c m^\m + \c^* \bar m^\n \ ,
\label{gb}
}
and let 
\EQ{
M^\m ~=~ A \ell^\m + B n^\m + C m^\m + D \bar m^\m \ .
\label{gc}
}
The null conditions imply $\a\b - \c\c^* = 0, ~AB-CD = 0$ and
$A\b + B\a - C\c^* - D\c = 0$.
The Weyl tensor for a Petrov type D metric can be written as 
(see ref.\cite{Chandra}, chapt.~1, eq.~(298), but note a typographical 
error)
\EQ{
C_{\m\n\l\r} ~~=~~\Psi_2\Bigl(-\frac12[\ell n \ell n] 
-\frac12[m \bar m m \bar m] + [\ell n m \bar m] + 
[\ell m n \bar m]\Bigr) ~~+~~ \text{c.c.} ~,
\label{gd}
}
where the notation $[ ~\ldots ~]$ indicates the linear combination of
the given vectors with the symmetries of the Weyl tensor.
A straightforward calculation now shows\footnote{
By direct calculation we find
$$
C_{LMLM} = -6 \Bigl( (A\b - C\c^*)^2 ~\Psi_2 ~+~ 
(A\b - D\c)^2 ~\Psi_2^* \Bigr)
$$
However, the null conditions can be manipulated to show that
$(A\b - C\c^*)(A\b - D \c) =0$, proving that at least one of the 
coefficients above vanishes. This establishes eq.\eqref{ge}.
In the Schwarzschild example below, we see explicitly that
$(A\b - C\c^*)$ is non-zero and gives $K_s$, while $(A\b - D \c)$
vanishes as required.}
\EQ{
C_{LMLM} ~=~ {3\over2} ~K_s^2 ~\Psi_2^{5\over3} \ ,
\label{ge}
}
where the complex quantity $K_s$ is defined as 
\SP{
K_s ~&=~ 2 \Psi_2^{-{1\over3}} (L\cdot\ell M\cdot n - L\cdot m M\cdot\bar m) \\
&=~ 2 \Psi_2^{-{1\over3}} (A\b - C \c^*) \ .
\label{gf}
}

The significance of this result follows from a theorem of Walker and
Penrose \cite{Walker}, stated and proved in ref.\cite{Chandra}
(theorem 1, chapter 7), that $K_s$ is conserved along the null
geodesic with tangent $L^\m$, {\it i.e.}
\EQ{
L^\m \nabla_\m K_s ~=~0 \ .
\label{gh}
}
This shows very directly how the Penrose limit is determined not just
by the curvature but also by an integral of motion along the chosen 
geodesic $\c$. We shall see explicitly how this is realized in the 
examples below.

The plane wave profile for the Penrose limit, allowing for a spacetime
with a non-vanishing Ricci as well as Weyl tensor, is
\EQ{
h_{ij} ~=~ {1\over2} \left(\begin{matrix}
(R_{LMLM} + R_{L\bar M L \bar M} + 2 R_{LML\bar M})
&~~-i(R_{LMLM} - R_{L\bar M L \bar M}) \\
-i(R_{LMLM} - R_{L\bar M L \bar M})
&~~ -(R_{LMLM} - R_{L\bar M L \bar M}- 2 R_{LML\bar M})\\
\end{matrix}\right) \ .
\label{gg}
}
From the definition of the Weyl tensor,
\SP{
R_{\m\n\l\r} ~=~ C_{\m\n\l\r} &+ 
{1\over2}\bigl(g_{\m\l}R_{\n\r} - g_{\m\r}R_{\n\l} - g_{\n\l}R_{\m\r}
+ g_{\n\r} R_{\m\l}\bigr) \\
&-{1\over6}\bigl(g_{\m\l}g_{\n\r} - g_{\m\r}g_{\n\l}\bigr)
\label{gggg}
}
and using the identity $C_{LML\bar M} = 0$, we immediately find
\EQ{
R_{LMLM} ~=~ C_{LMLM}, ~~~~~~~~~~R_{LML\bar M} ~=~ {1\over 2} R_{LL}\ .
\label{gghh}
}
The eigenvalues of $h_{ij}$ are therefore
\SP{
h_{\pm} ~&=~ {1\over2} R_{LL} ~\pm~ |C_{LMLM}| \\
&=~ - \hat\Phi_{00} ~\pm~ |\hat\Psi_{0}| \ ,
\label{ggii}
}
the latter form emphasizing again the importance of the NP scalars
in the description of the Penrose limit and photon 
propagation\cite{Shore:1995fz}.

Using \eqref{ge}, we have therefore shown that the Penrose limit
for a general type D spacetime has a profile function $h_{ij}$ with
eigenvalues
\EQ{
h_{\pm} ~=~ 
{1\over 2}R_{LL} ~\pm~ {3\over2}|K_s|^2 |\Psi_2|^{5\over3} \ .
\label{ggjj}
}
Note that, as expected, $h_{ij}$ is traceless when the original metric 
is Ricci flat.

In fact, this result can be simplified still further. Following
Chandrasekhar \cite{Chandra}, we can prove using the orthogonality
and null properties of the $L,N,M,\bar M$ basis (including the result
that if $A\b - C\c^* \neq 0$ then $A\b - D\c = 0$; see footnote 15),
that
\EQ{
|K_s|^2 ~=~ 4 |\Psi_2|^{-{2\over3}} \a \b \ .
\label{ggkk}
}
So we can finally write
\EQ{
h_\pm ~=~ {1\over2}R_{LL} ~\pm~ 6 |\Psi_2| L\cdot \ell L\cdot n \ .
\label{ggll}
}
This shows very clearly that in the case of Type D spacetimes, the
Penrose limit depends only on the tangent vector 
$L^\m \equiv \hat k^\m$ itself; we do not even need to compute
the remaining basis vectors in the set $E_A$.

The advantages of this method of determining the Penrose limit are now clear: it only involves knowledge of the classical null geodesic $\c$ itself and does not invoke any particular embedding into a null 
congruence; it makes the relation with the conserved quantities characterizing the classical orbit manifest; and it allows the 
simplifying power of the NP formalism and Petrov classification 
to be exploited fully.

\subsection{Schwarzschild black hole}

The Schwarzschild metric is
\EQ{
ds^2 ~=~ -{\D\over r^2}dt^2 + {r^2\over\D}dr^2 + r^2(d\theta^2 + 
\sin^2\theta d\phi^2)
\label{gi}
}
where $\D = r^2 - 2Mr$. We consider planar classical orbits, and
with no loss of generality choose $\phi = {\rm const.}$ The first 
integrals of the null geodesic equations give
\EQ{
\dot t = E {r^2\over\D}, ~~~~~~~~
\dot r = E \sqrt{1 - {b^2\D\over r^4}}, ~~~~~~~~
\dot \theta = E {b\over r^2}, ~~~~~~~~
\dot\phi = 0 \ ,
\label{gj}
}
where $E$ is the energy and $b = L/E$ is the impact parameter, 
with $L$ the conserved angular momentum.
We can therefore identify the tangent vector $\hat k^\m$ as
$\bigl({r^2\over\D}, F, {b\over r^2}, 0\bigr)$ in the 
$(t,r,\theta,\phi)$ coordinate basis, where  
$F= \sqrt{1- b^2\D/r^4}$.

To derive the Penrose limit, we first identify a parallel transported
frame $E_A$ along this null geodesic, as described above. 
We immediately have 
\EQ{
E_u{}^\m ~\equiv~ \hat k^\m ~=~ \bigl({r^2\over\D}, 
F, {b\over r^2}, 0\bigr) \ .
\label{gk}
} 
The transverse vierbeins $E_i^\m$ are then determined by the 
orthonormality conditions $g_{\m\n}E_A{}^\m E_B{}^\n = \eta_{AB}$, 
with $\eta_{AB}$ as implied in \eqref{cl}, supplemented by the parallel transport condition $\hat k^\m \nabla_\m E_i{}^\n = 0$. The 
orthonormality conditions give
\EQ{
E_1{}^\m ~=~ {1\over r}\Bigl({r^4\over b\D}(c-F),~ {r^2\over b}(cF-1),~
c,~0\Bigr) \ ,
\label{gl}
}
and
\EQ{
E_2{}^\m ~=~ {1\over r\sin\theta} \bigl(0,~0,~0,~1\bigr) \ ,
\label{gm}
}
showing explicitly that these algebraic conditions determine 
$E_1{}^\m$ only up to the addition of a piece proportional to 
$\hat k^\m$ itself. The remaining freedom, parameterized by the function
$c(r)$ above, is determined by the parallel transport condition, which
gives $c(r) = {1\over r} \int dr~F^{-1}$. Note that we do not 
require an explicit expression for $E_v{}^\m$ for this construction.

The next step is to determine the matrix $\W_{ij}$,
defined as
\EQ{
\W_{ij} ~=~ E_i{}^\m E_j{}^\n \nabla_\m \hat k_\n \ .
\label{gn}
}
This calculation can be simplified by noticing that since
$\hat k^\m \nabla_\m \hat k^\n = 0$ by the geodesic equation,
and $\nabla_\m \hat k_\n$ is symmetric because $\hat k^\m$ is a 
gradient flow, we can add a multiple of $\hat k^\m$ to the vierbeins
$E_i{}^\m$ in \eqref{gn} without affecting the result for $\W_{ij}$.
In practice, this allows us to choose the function $c(r)$ in \eqref{gl}
as we like in order to make the subsequent calculation simple.
Natural choices include $c = F$ \cite{Shore:2007um} and 
$c = F^{-1}$ \cite{Blau2}. Exploiting this freedom, we find after
a straightforward calculation that $\W_{ij}$ is diagonal in this basis
with
\EQ{
\W_{11} ~=~ {1\over rF} \Bigl( 1 - {M b^2\over r^3}\Bigr), ~~~~~~~~~~
\W_{22} ~=~ {F\over r} + {b\over r^2} \cot\theta \ .
\label{go}
}
Alternatively, using the geodesic equations \eqref{gj}, we can express
these in the more enlightening form:
\EQ{
\W_{11} ~=~ {d\over du} \log(r \dot r), ~~~~~~~~~~
\W_{22} ~=~ {d\over du} \log(r\sin\theta) \ .
\label{gp}
}
These results are clearly consistent with the trace identity \cite{Blau2}
\EQ{
{\rm tr}~\W ~=~ \eta^{AB} E_A{}^\m E_B{}^\n \nabla_\m \hat k_\n
~=~ \nabla_\m \hat k^\m ~=~ 
{1\over\sqrt{g}} \partial_\m \bigl(\sqrt{g} \hat k^\m\bigr) \ .
\label{gq}
}
The profile function $h_{ij} = \dot\W_{ij} + \W_{ik}\W^k{}_j$ 
for the plane wave is easily computed from \eqref{gp} and we find
\EQ{
h_{11} ~=~  -{1\over{r \dot r}} {d^2\over du^2} (r \dot r), ~~~~~~~~~~
h_{22} ~=~  -{1\over{r \sin\theta}} {d^2\over du^2} (r \sin\theta) \ ,
\label{grr}
}
that is,
\EQ{
h_{11} ~=~  -{3M b^2 \over r^5}, ~~~~~~~~~~
h_{22} ~=~ {3M b^2 \over r^5} \ .
\label{gr}
}
As anticipated, ${\rm tr}~h_{ij} = 0$ as the Penrose limit inherits
the Ricci flat property of the original Schwarzschild spacetime.

In the Newman-Penrose method, we first note that the NP null tetrad
for the Schwarzschild metric is\cite{Chandra,Shore:2007um}:
\EQ{
\ell^\m = {1\over\D}(r^2,\D,0,0),~~~~~~~~
n^\m = {1\over 2r^2}(r^2,-\D,0,0),~~~~~~~~
m^\m = {1\over{\sqrt{2}r}}(0,0,1,{i\over\sin\theta})\ ,
\label{gs}
}
and the only non-vanishing component of the Weyl tensor is 
$\Psi_2 = -M/r^3$.
We can therefore expand the tangent vector $L^\m \equiv \hat k^\m$ for
the classical orbit in the NP basis as follows:
\EQ{
L^\m ~=~ {1\over2}(1+ F)\ell^\m + {r^2\over\D}(1-F)n^\m 
+ {b\over{\sqrt{2} r}}(m^\m + \bar m^\m) \ .
\label{gt}
}
The circular polarization vector $M^\m$,
which satisfies $L.M = 0, ~M.\bar M=1$ and is parallel transported, 
is identified with the vierbein found above:
\SP{
M^\m ~&=~ {1\over\sqrt2} \bigl(E_1{}^\m + i E_2{}^\m\bigr) \\
&=~ {1\over{\sqrt{2} r}}
\Bigl({r^4\over b\D}(c-F),~ {r^2\over b}(cF-1),~ c,~ 
{i\over{\sin\theta}}\Bigr) \ .
\label{gu}
}
Expanding this in the standard NP basis we have
\EQ{
M^\m ~=~ {1\over{\sqrt{2} r}}{r^2\over b} 
\Bigl(-{1\over 2}(1-c)(1+F)\ell^\m + {r^2\over \D}(1+c)(1-F)n^\m\Bigr)
+ {1\over2}(1+c) m^\m - {1\over2}(1-c) \bar m^\m \ .
\label{gv}
}
This fixes the coefficients $\a,\ldots,\c$ and $A,\ldots,D$ of 
eqs.\eqref{gb} and \eqref{gc}. Once again we can
use the fact that adding a multiple of $L^\m$ to $M^\m$ affects
neither the evaluation of $K_s$ (since $L^\m$ is null) or $C_{LMLM}$.
So to perform the calculation efficiently, we can make any convenient
simplifying choice of $c(r)$ in \eqref{gv}. From \eqref{gf} we then
readily find the conserved Walker-Penrose quantity $K_s$:
\SP{
K_s ~&=~ -\sqrt{2} (-M)^{-{1\over3}} b \ .
\label{gw}
}
We see directly that in this simple case the conserved quantity $K_s$
for a given photon energy is determined by the angular momentum of 
the orbit. This confirms the advertised link between the conservation
laws governing the classical null geodesic and the Penrose limit.
The Penrose plane wave profile itself \eqref{gr} follows immediately.
 
\vskip0.2cm
We can make a number of immediate observations based on these results 
(see also ref.\cite{Blau2}):

\noindent(i)~~The explicit dependence of $h_{ij}$ on the impact
parameter $b$ implies that $h_{ij}=0$ for classical null trajectories 
with vanishing angular momentum. This confirms that the Penrose limit 
is flat, and therefore the refractive index receives no quantum corrections, for purely radial geodesics which of course are the 
principal null geodesics of the Schwarzschild metric.

\noindent(ii)~~There is a closed photon orbit at a critical value 
$r_c$ of the radial coordinate $r$. Using \eqref{gj} we have 
$\ddot r = E {b^2\over r^3}\bigl(1 - {3M\over r}\bigr)$,
fixing $r_c = 3M$.  Then, requiring $\dot r = 0$ determines 
the critical impact parameter $b_c = 3\sqrt{3}M$.  
For this orbit, the Penrose limit has
\EQ{
h_{11}~=~ - h_{22} ~=~ -{3M b_c^2\over r_c^5} ~=~ 
-{1\over 3M^2}\ ,
\label{gga}
}
which describes a Ricci flat symmetric plane wave. The refractive index 
for this case has already been discussed in section 6.

\noindent(iii)~~ In the near-singularity limit, the radial geodesic equation becomes
\EQ{
{\dot r}^2 ~=~ {2Mb^2\over r^3} ~+~ O\Bigl({1\over r^2}\Bigr) \ ,
\label{ggb}
}
with solution $r^5 = {25\over2}Mb^2 u^2$.
The Penrose limit therefore has
\EQ{
h_{11}~=~ - h_{22} ~=~ -{6\over25} {1\over u^2}\ .
\label{ggc}
}
Note that this is independent of the impact parameter $b$, though 
remember the derivation is only valid for orbits with $b\neq 0$. Also, 
in order for an incoming photon to reach the singularity, the impact parameter must be sufficiently small, $b < b_{\rm crit} = 3\sqrt{3}M$.
The Penrose limit in the near-singularity region of Schwarzschild 
spacetime is therefore a Ricci flat singular homogeneous plane wave, 
with coefficients $c^{1,2} = \mp 6/25$. As explained in detail in 
\cite{Blau2,Blau:2003dz,Blau:2004yi}, this specific coefficient is 
related to the power-law nature of the singularity for this class of 
black hole spacetimes.

\vskip0.2cm
Having established the Penrose limit, we now need to determine the 
Van Vleck-Morette matrix $\BDelta(u,u')$. This entails solving the 
Jacobi equation for $\BA(u,u'$ with boundary conditions $\BA(u',u') = 0$, ${d\over du}\BA(u,u')|_{u=u'} = \Bone$. Recall that the Jacobi equation 
for the geodesic deviation vector $y^i$ in Brinkmann coordinates
is in general
\EQ{
{d^2\over du^2}y^i+h_{ij} y^j=0 \ ,
\label{ggd}
}
and in the present case, $h_{ij}$ is diagonal. Beginning with $i=1$, 
note that the special form \eqref{grr} of the expression for $h_{11}$ 
shows immediately that a particular solution is 
$y^1 = f_1(u) = r(u) \dot r(u)$ with $r(u)$ the solution of the 
geodesic equation \eqref{grr}. 
To find the general solution, recall that given two solutions of this second order ODE, the Wronskian $W = f_1 \dot f_2 - \dot f_1 f_2$ is a constant. It follows that the general solution can be written in the 
form
\EQ{
y^1(u) ~=~ c_1 f_1(u) \int_{c_2}^u {du\over f_1(u)^2} \ .
\label{gge}
}
Imposing the relevant boundary conditions, we deduce
\EQ{
A_{11}(u,u') ~=~  f_1(u) f_1(u') \int_{u'}^u {du\over f_1(u)^2} \ .
\label{ggf}
}
In the same way, a particular solution for $i=2$ is given from \eqref{grr} 
as $y^2 = g_1(u) = r(u) \sin\theta(u)$. in this case, it is easy to see 
by inspection (in fact, as a consequence of angular momentum conservation) that the general solution can be written simply in the form
\EQ{
y^2(u) ~=~c_1 r(u) \sin(\theta(u) + c_2)\ .
\label{ggg}
}
This direction is obviously the focusing one since
$y^2(u)$ oscillates with a amplitude that decreases as the
singularity is approached. The other direction de-focuses and $y^1(u)$
increases as the singularity is approached.
Applying the boundary conditions, we find
\EQ{
A_{22}(u,u') ~=~ {1\over b} r(u) r(u') \sin\bigl(\theta(u) - \theta(u')\bigr)\ .
\label{ggh}
}
This determines the VVM matrix $\BDelta(u,u')$. The non-vanishing 
components are
\EQ{
\D_{11}(u,u') ~=~ (u-u') \Bigl[r(u) \dot r(u) r(u') \dot r(u') 
\int_{u'}^u {du\over (r \dot r)^2}\Bigr]^{-1} \ ,
\label{ggi}
}
and 
\EQ{
\D_{22}(u,u') ~=~ b(u-u')
\Bigl[r(u)r(u')\sin \Bigl(b\int_{u'}^u {du\over r^2}\Bigr)\Bigr]^{-1} \ .
\label{ggj}
}

With these expressions for the VVM matrix, the refractive index for 
scalar and spinor QED can now be evaluated by substituting into the formulae \eqref{dj}, \eqref{eac} using the explicit solution $r(u)$ of 
the geodesic equation. The solution, however, does not show any special algebraic simplicity and an explicit numerical evaluation is not expected 
to reveal any distinctive new physics beyond what can be deduced by considering the particular limits and special cases discussed above. 

Rather more enlightening is the study of the refractive index and its analytic properties in the near-singularity limit. As shown above, the Penrose limit in this region is a Ricci flat singular homogeneous plane wave with $h_{11} = -h_{22} = -{6\over25}{1\over u^2}$.
The VVM matrix components are therefore given by substituting the
coefficients 
$\alpha_1=\tfrac75$ and $\alpha_2=\tfrac15$ 
into the general formula \eqref{ffc}. This gives
\EQ{
\D_{11}(u,u') ~=~ \frac{7(uu')^{1/5}(u-u')}{5(u^{7/5}-u^{\prime7/5})}
\ ,\qquad
\D_{22}(u,u') ~=~ \frac{u-u'}{5(uu')^{2/5}(u^{1/5}-u^{\prime1/5})} \ .
\label{ggl}
}

\begin{figure}[ht] 
\centerline{\includegraphics[width=2.5in]{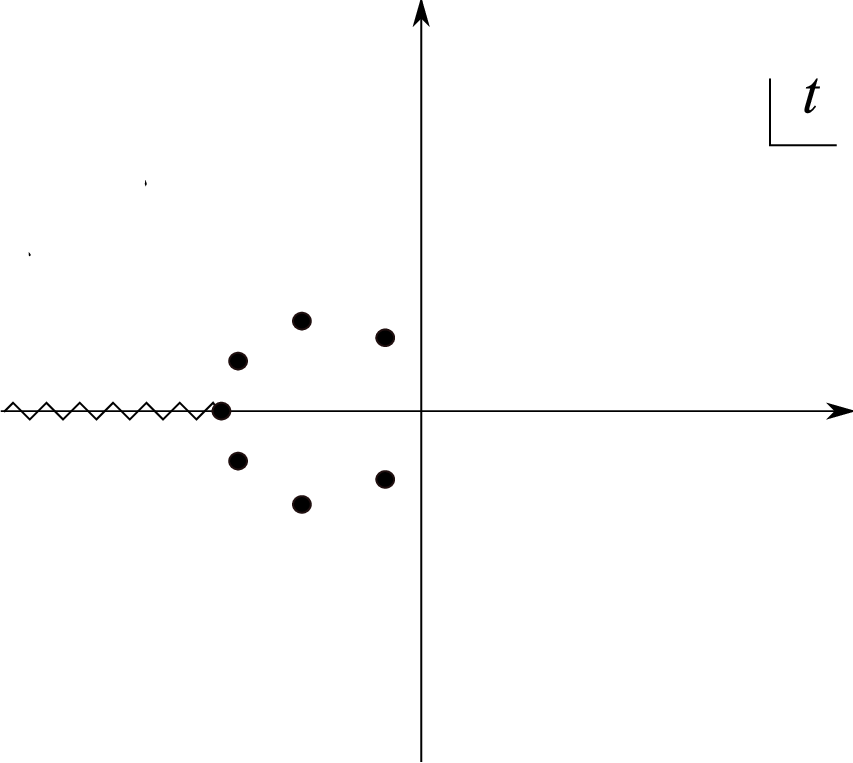}}
\caption{\footnotesize Analytic structure of the $t$ integrand 
of ${\cal F}^\text{scalar}(u;z)$ 
in the near-singularity region of Schwarzschild
spacetime. The branch point lies at the position
of the singularity and the other points are double poles.}
\label{f9}
\end{figure}
For scalar QED, the refractive index is given by \eqref{ffd}
with the other polarization obtained by swapping the powers
$\tfrac32\leftrightarrow\tfrac12$. 
Notice that the affine parameter $u$ is negative, with the 
singularity at $u=0$. In this case,
the integrand has a branch point at $t=-|u|$, {\it i.e.\/}~when $u'$ lies 
at the singularity, and double poles at
\EQ{
t=|u|\big(e^{2\pi ip/7}-1\big)\ ,\qquad p=1,2,\ldots,6\ .
}
as shown in Fig.\ref{f9}.

\begin{figure}[ht]
\centerline{\includegraphics[width=2.5in]{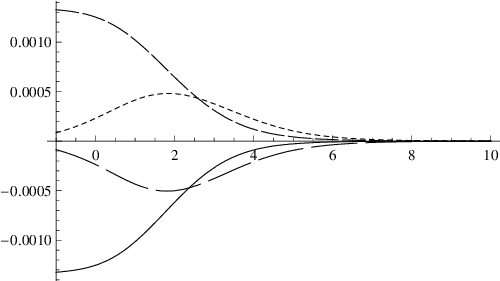}
\hspace{0.2cm}\includegraphics[width=2.5in]{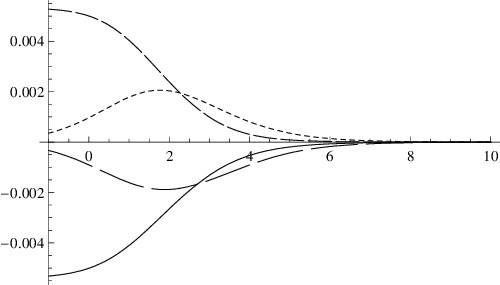}}
\caption{\footnotesize $n(\omega)-1$ for the near singularity region of
  the Schwarzschild black hole
  plotted as a function of $\log\omega$:
  continuous (real part, polarization $i=1$); big dashes (imaginary
  part, $i=1$); small dashes 
(real part, polarization $i=2$) and dots (imaginary
  part, $i=2$), for scalar QED (left) and QED (right).}
\label{f5}
\end{figure}
The refractive index in the near singularity limit can be evaluated
numerically and the results are shown in Fig.\ref{f5}. As expected
for a Ricci flat spacetime, the signs of the corrections are the
same for scalar and spinor QED, with both showing birefringence.
In this example, however, unlike the case of the Ricci flat symmetric
plane wave, the sign of $\IM n(\w)$ differs for the different 
polarization states.

\vskip1cm

\subsection{Kerr black hole}

The case of the Kerr spacetime describing a rotating black hole can
be analyzed in the same way and provides a very nice illustration of
the simplicity of the Newman-Penrose tetrad approach. 
The Kerr metric is: 
\EQ{
ds^2 ~~=~~ - \r^2 {\D\over \S^2} dt^2 ~+~ \r^2 {1\over \D} dr^2
~+~ \r^2 d\theta^2 ~+~ {1\over \r^2}\S^2 \sin^2\theta \bigl(
d\phi - \w dt\bigr)^2 \ ,
\label{ggga}
}
where $\D = r^2 - 2Mr + a^2$, ~$\S^2 = (r^2+a^2)^2 - a^2\sin^2\theta\D$,
~$\r^2 = r^2 + a^2 \cos^2\theta$ and $\w = 2Mra/\S^2$.
The metric is specified by two parameters, $M$ and $a$, where $M$ is the
mass and $Ma$ the angular momentum (about the rotation axis $\theta = 0$)
as measured from infinity. Note that the time component simplifies,
$g_{tt} = (1-2Mr/\r^2)$.

The condition $\D(r) = 0$, for which there is a coordinate singularity
similar to that in the Schwarzschild metric, has two real solutions,
$r = r_{\pm} = M \pm \sqrt{M^2 - a^2}$ (assuming $a<M$). The larger, 
$r=r_+$, is the event horizon. The region $r <  r_-$ contains a ring singularity. However, for the Kerr metric, this does not coincide with
the condition $g_{tt} = 0$, which defines the boundary of the region,
the ergosphere, where an asymptotically timelike Killing vector 
becomes spacelike. Within the ergosphere, even null curves are pulled 
round in the direction of the rotation. Its outer limit
$r_E(\theta) = M +\sqrt{M^2 - a^2 \cos^2\theta}$, known as the 
stationary limit surface, coincides with the event horizon at the poles 
and is equal to the Schwarzschild radius $2M$ at the equator.

The null geodesics in the Kerr metric have been extensively studied
(see ref.\cite{Chandra} for a comprehensive discussion, and
\cite{Daniels:1995yw} for a first analysis of photon propagation
in QED). Although the
equatorial orbits are of special interest and simplicity, our formalism
allows us to consider immediately the general case of non-planar orbits.
In this case, analogous to \eqref{gj}, we have 
$\hat k^\m = E^{-1}(\dot t, \dot r, \dot\theta, \dot\phi)$ where
\SP{
\dot t ~&=~ {E\over\D} \Bigl(r^2 + a^2 + 
{2Mra\over\r^2}(a\sin^2\theta -b)\Bigr) \\
\dot r ~&=~ {E\over\r^2} \Bigl(r^4 + (a^2 - b^2 - q)r^2
+ 2Mr \bigl((a-b)^2 + q\bigr) - a^2 q\Bigr)^{1\over2} \\
\dot\theta ~&=~ {E\over\r^2} \Bigl(q + (a^2\sin^2\theta - b^2)
\cot^2\theta \Bigr)^{1\over2} \\
\dot\phi ~&=~ {E\over \D \sin^2\theta}\Bigl(b + {2Mr\over\r^2}
(a\sin^2\theta - b)\Bigr) \ .
\label{gggb}
}
The orbits are characterized by the energy $E$ and two impact parameters,
the familiar $b = L_z/E$ and $q$, its analogue in the $\theta$-plane
\cite{Chandra}. Both are conserved quantities.

Like Schwarzschild, the Kerr spacetime is Petrov type D and the only
non-vanishing component of the Weyl tensor is 
$\Psi_2 = C_{\ell m n \bar m}$ referred to the null basis\footnote{
The equivalent covariant vectors required for the calculations here are
$$
\ell_\m= \bigl(-1,{\r^2\over\D},0,a\sin^2\theta\bigr) \ ,~~~~~~~~~~~
n_\m = {\D\over2\r^2}\bigl(-1, -{\r^2\over\D},0,a\sin^2\theta\bigr)
$$
}
\SP{
\ell^\m ~&=~ {1\over\D}\bigl(r^2 + a^2, \D, 0, a\bigr) \\
n^\m ~&=~ {1\over2\r^2}\bigl(r^2 + a^2, -\D, 0, a\bigr) \\
m^\m ~&=~ {1\over\sqrt{2} \tilde \r}
\bigl(i a \sin\theta, 0, 1, {i\over\sin\theta}\bigr) \ ,
\label{gggc}
}
where $\tilde\r = r + i a\cos\theta$. In this basis, 
$\Psi_2 = -{M / ({\tilde\r}^*)^3}$.

At this point, we can simply invoke the general discussion of type D
spacetimes in section 7.1 to write down the Penrose limit. The Kerr
spacetime is Ricci flat, so from \eqref{ggll} we see that the 
eigenvalues of the plane wave profile function $h_{ij}$ corresponding 
to the null geodesic with tangent vector $\hat k^\m$ are simply
\EQ{
h_{ij} ~=~ \delta_{ij}(-1)^i|\Psi_2|~ \hat k\cdot \ell~ \hat k\cdot n \ ,
\label{gggd}
}
with $|\Psi_2| = M/\r^3$. A straightforward algebraic exercise now 
shows that 
\EQ{
\hat k\cdot \ell ~\hat k\cdot n ~=~ {1\over 2\r^2} \bigl((a-b)^2 + q\bigr) \ ,
\label{ggge}
}
consistent with the general theorem that
\EQ{
|K_s|^2~=~ 4|\Psi_2|^{-{2\over3}}~ \hat k\cdot\ell ~\hat k\cdot n 
~=~ 2 M^{-{2\over3}} \bigl((a-b)^2 + q\bigr) 
\label{gggf}
}
is conserved along the geodesic. For the general Kerr orbit, we
therefore have 
\EQ{
h_{11} ~=~ -{3M\over\r^5} \bigl((a-b)^2 + q\bigr) \ ,\qquad
h_{22} ~=~ {3M\over\r^5} \bigl((a-b)^2 + q\bigr)
\label{gggg2}
}
This is a remarkable simplification and illustrates very clearly the 
dependence of the Penrose limit on both the spacetime curvature
and the characteristics of the chosen null geodesic.

Eq.\eqref{gggg2} is a straightforward generalization of the 
Schwarzschild result \eqref{gr} and similar consequences follow here.
Clearly, the Penrose limit is flat for a principal null geodesic,
{\it i.e.}~$\hat k^\m = \ell^\m$. Once again, there is an unstable
closed orbit for a constant value $r_c$ of the radial
coordinate. Imposing $\ddot r = 0$, and specializing for simplicity
to equatorial orbits, determines the relation 
$r_c = 3M (b_c - a )/(b_c + a)$,
though here the expression for the critical radius is more 
complicated, involving the solution of a cubic equation, and we have
$r_c = 2M\Bigl[1 + \cos\Bigl({2\over3}\cos^{-1}\bigl(
\pm{a\over M}\bigr)\Bigr)\Bigr]$ 
for the retrograde and direct orbits respectively \cite{Chandra}.
Evaluating $h_{ij}$ for this orbit, we find the non-vanishing elements
\EQ{
h_{11} ~=~ -{3\over r_c^2} ~=~
-{1\over3M^2} \Bigl({b_c +a\over b_c -a}\Bigr)^2 \ ,
\qquad
h_{22} ~=~ {3\over r_c^2} ~=~
{1\over3M^2} \Bigl({b_c +a\over b_c -a}\Bigr)^2
\label{gggh}
}
Since this is a constant, we confirm that, just as in the 
Schwarzschild case, the Penrose limit for the critical orbit in 
Kerr spacetime is a Ricci flat symmetric plane wave.

The near-singularity limit follows straightforwardly. Restricting 
to equatorial orbits, we determine the small $r$ behaviour of the
radial geodesic equation from \eqref{gggb} as 
\EQ{
{\dot r}^2 ~=~ {2M(a-b)^2\over r^3} + 
{\cal O}\Bigl({1\over r^2}\Bigr) \ ,
\label{gggi}
}
and integrating, we find $r^5 = {25\over2}M(b-a)^2 u^2$ and therefore 
the Penrose limit profile function is a homogeneous plane wave identical
to the Schwarzschild result \eqref{ggc}. This reflects the
equivalent power-law 
nature of the singularities for Schwarzschild and Kerr black holes.
This result implies that the singularity behaviour of the refractive
index is the same as for the Schwarzschild solution.

\vskip0.5cm

\subsection{Reissner-Nordstrom black hole}

It is very simple to generalize the calculation of the Penrose Limit
for the Schwarzschild black hole to the Reissner-Nordstrom black
hole. Here we shall simply state the result that
\EQ{
h_{11}=-\frac{3Mb^2}{r^5}+\frac{4b^2Q^2}{r^6}\ ,\qquad
h_{22}=\frac{3Mb^2}{r^5}-\frac{2b^2Q^2}{r^6}\ .
}
Notice that the hole is not Ricci flat so that
\EQ{
R_{uu}=h_{11}+h_{22}=\frac{2b^2Q^2}{r^6}\ .
}

This result is not as it stands useful for the calculation of the
refractive index because in this case there is background
electro-magnetic field that would have to be taken into account in
addition to the curvature of spacetime \cite{Daniels:1993yi}.

\section{Cosmological spacetimes}
\label{cosmo}

We now consider a cosmological spacetime with an initial singularity,
the Friedmann-Robertson-Walker universe. This example is 
of more than just conceptual interest since modifications to the speed 
of light in the very early universe are potentially important in the context of the horizon problem, one of the original motivations for inflation.

The metric for a FRW spacetime has the form
\EQ{
ds^2 ~=~ -dt^2+a(t)^2\Big[dr^2+f_\kappa(r)d\Omega^2\Big]
\label{ia}
}
where $f_\kappa(r)=r$, $\sin r$ and $\sinh r$ for $\kappa=0$, $+1$ and
$-1$ respectively, representing a flat, closed or open universe, and
$d\Omega^2$ is the usual metric for the two-sphere. With no loss
of generality, we consider geodesics with zero angular momentum on
this transverse space. These null geodesics are well known.
We have
\EQ{
\dot t=\frac1a\ ,\qquad \dot r=\frac1{a^2}\ ,
\label{ib}
}
Note that we use the notation $\dot t\equiv {dt\over du}$ for 
$u$-derivatives; in what follows we denote $t$-derivatives by 
$a'(t) \equiv {da\over dt}$, {\it etc\/}. Here, $u$ is as usual the
affine parameter along the null geodesic $\c$. We therefore find
$\hat k^\m = \bigl(a^{-1},a^{-2},0,0\bigr)$ in a $(t,r,\theta,\phi)$
coordinate basis.

The first step in finding the Penrose limit by the conventional
method \cite{Blau2} described earlier is to identify a parallel
transported frame $E_A$ along $\c$. Here, we have:
\SP{
E_u{}^\m ~\equiv~ \hat k^\m ~&=~ \bigl(a^{-1}, a^{-2}, 0, 0\bigr) 
~~~~~~~~~~~~~~
E_v{}^\m ~=~ -{1\over2}\bigl(-a, 1, 0, 0\bigr) \\
E_1{}^\m ~&=~ (af_\k)^{-1}\bigl(0, 0, 1, 0\bigr) 
~~~~~~~~~~
E_2{}^\m ~=~ (af_\k)^{-1}\bigl(0, 0, 0, 1/\sin\theta\bigr) 
\label{ic}
}
The matrix $\Omega_{ij} = E_i{}^\m E_j{}^\n \nabla_\m \hat k_\n$~
$(i,j = 1,2)$~ follows straightforwardly and we find after a short
calculation that\footnote{The calculations in this section involve 
the following Christoffel symbols:
$$
\Gamma^t_{tr} = a a' ~~~~~~
\Gamma^r_{rt} = a^{-1}a'
$$
$$
\Gamma^t_{\theta\theta} = aa' f_\k^2 ~~~~~~
\Gamma^t_{\phi\phi} = aa' f_\k^2 \sin^2\theta ~~~~~~
\Gamma^r_{\theta\theta} = - f {\partial f\over\partial r} ~~~~~~
\Gamma^r_{\phi\phi} = - f {\partial f\over\partial r} \sin^2\theta \  .
$$}
\EQ{
\Omega_{ij} ~=~ \d_{ij} {d\over du} \log(af_\k) \ ,
\label{id}
}
with $a \equiv a(t(u))$, $f_\k \equiv f_\k(r(u))$ along $\c$.
The profile function for the Penrose plane wave is then
\EQ{
h_{ij} ~=~ -\d_{ij} (a f_\k)^{-1} {d^2\over du^2}(a f_\k) \ .
\label{ie}
}
This confirms $h_{11} = h_{22}$, as expected since the Penrose limit 
inherits the conformally flat property of the original FRW metric.
Then, since the metric function $f_\k(r)$ satisfies
${\partial^2\over\partial r^2}f_\k + \k f_\k = 0$ for all $\k$, 
we readily find
\cite{Blau2}
\EQ{
h_{ij} ~=~ -\d_{ij} \Bigl({\ddot a\over a} - {\k\over a^4}\Bigr) \ .
\label{if}
}
Finally, to make contact with conventional FRW dynamics, we can
re-express this in terms of $t$-derivatives, giving
\SP{
h_{ij} ~&=~ -\d_{ij} {1\over a^2} \Bigl({a''\over a} - 
{\k + a^{\prime 2} \over a^2}\Bigr) \\
&=~ \d_{ij} {4\pi G\over a^2} \r (1 + w) \ , 
\label{ig}
}
using the standard Friedmann and acceleration equations,
\SP{
{a^{\prime 2}\over a^2} + {\k\over a^2} ~&=~ {8\pi G\over 3} \r \\
{a^{\prime \prime}\over a} ~&=~ - {4\pi G\over3} (\r + 3p) 
\label{ih}
}
and the equation of state $p = w\r$.

\vskip0.2cm
The Newman-Penrose null tetrad derivation of the Penrose limit 
is especially quick in this example. Recall from the discussion in
the preceding section that the profile function is given by \eqref{gg}, where here we can identify the NP basis vectors $L, N, M, \bar M$ 
with the $E_A{}$ of \eqref{ic}. Since the FRW spacetime is conformally 
flat, we immediately have
\EQ{
h_{ij} ~=~ {1\over 2} R_{\m\n}E_u{}^\m E_u{}^\n \d_{ij} ~\equiv~ 
{1\over 2}R_{uu} \d_{ij} \ .
\label{ii}
}
To check this with the previous result \eqref{ig}, recall that the 
Ricci tensor is given by 
$R_{\m\n} = 8\pi G (T_{\m\n} - {1\over2}T^\l{}_\l g_{\m\n})$
where the FRW energy-momentum tensor is
$T_{\m\n} = (\r + p) t_\m t_\n + p g_{\m\n}$ with $t^\m$ a unit vector
in the $t$-direction.
It follows immediately that
\SP{
h_{ij} ~&=~ {1\over 2} R_{\m\n} E_u{}^\m E_u{}^\n \d_{ij} \\
&=~ 8\pi G (\r + p) (E_u \cdot t)^2 \\
&=~ \delta_{ij}{4\pi G\over a^2} (\r + p)\ ,  
\label{ij}
}
since the terms proportional to $g_{\m\n}$ do not contribute as
$E_u{}^\m$ is null.

\vskip0.2cm
To take this further, we now use the standard conservation equation
\EQ{
{d\over da} \bigl(\r a^3\bigr) ~=~ - 3 w \r a^2 \ ,
\label{ik}
}
which follows from $\nabla_\m T^{\m\n} = 0$, to deduce as usual
$\r \sim a^{-3(1+w)}$. It is convenient here to introduce the parameter
$\c = {2\over 3(1+w)}$. The three standard equations of state considered
in FRW cosmology are $w=0 ~(\c = 2/3)$ for non-relativistic matter,
$w = 1/3 ~(\c = 1/2)$ for radiation, and $w = -1 ~(\c \rta \infty)$
for a cosmological constant. The latter implies a flat Penrose
limit, as is evident from \eqref{ig}, since a vacuum spacetime with a
cosmological constant, {\it i.e.\/}~de Sitter space, 
is maximally symmetric. Combining \eqref{ik} 
with the Friedmann equations, and introducing the constant
$C_\c {8\pi G\over3}\r a^{2\over\c}$, we now have
\SP{
\k + a^{\prime 2} ~&=~ C_\c a ^{2(1- {1\over\c})} \\
a^{\prime \prime} ~&=~ C_\c \bigl(1 - \tfrac1{\c}\bigr) a^{1 - {2\over\c}}
\label{il}
}
showing clearly that $\c < 1 ~(w < -1/3)$ is the threshold for an accelerating universe. We therefore find from \eqref{ig} that
the Penrose limit profile function is
\EQ{
h_{ij} ~=~ \d_{ij} C_\c \c^{-1}  a^{-2(1 + {1\over\c})} \ ,
\label{im}
}
where the dependence of $a(u)$ on $u$ follows from \eqref{il} as
\EQ{
u ~=~ \int da {a\over \sqrt{C_\c a^{2(1-{1\over\c})} - \k}} \ .
\label{in}
}

It is now simplest to consider the various possible cases separately.

\vskip0.1cm
\noindent(i)~~Spatially flat universe, $\k=0$, for any $\c$:

In this case, the integral in \eqref{in} is trivial and the dependence
of $a(u)$ on $u$ is given implicitly by $u = C_\c^{-1/2}
{\c\over\c+1}a^{1+\c^{-1}}$. The profile function \eqref{im} follows
immediately:
\EQ{h_{ij} ~=~ {\c\over(\c+1)^2} {1\over u^2} \d_{ij} \ .
\label{io}
}
Perhaps unsurprisingly given our experience with black hole spacetimes
with singularities in the previous section, we find here that the
Penrose limit of a flat FRW spacetime is a conformally flat singular
homogeneous plane wave with coefficient $c = {\c\over(\c+1)^2}$. 

\vskip0.2cm
\noindent(ii)~~$\c = 1$, for any $\k$:

As already noted, an equation of state with $\c = 1 ~(w = -{1\over3})$
is on the boundary between a decelerating ($0<\c<1$) and accelerating
($\c>1$) universe. From the Friedmann equation, we see this
requires $C_1 - \k > 0$, which only becomes a constraint on the energy
density for a closed universe, $\k = 1$. In this case, we find
$u = {1\over2} {1\over \sqrt{C_1 - \k}} a^2$, and the profile function
can be expressed as
\EQ{
h_{ij} ~=~ {1\over4} \Bigl(1 + {\k\over C_1 - \k}\Bigr) {1\over u^2}
\d_{ij} \ .
\label{ip}
}
Again we find a singular homogeneous plane wave, though with a
$\k$-dependent modification of the coefficient in \eqref{io}.
The case $\kappa=1$ is interesting in that in this case
$\alpha_i^2=(1-C_1)^{-1}<0$ and so $\alpha_i$ is imaginary and,
following the discussion in Section \ref{shpw}, there
are an infinite sequence of conjugate points.

In these special cases, the characteristic $1/u^2$ behaviour of
$h_{ij}$ holds exactly for all $u$, {\it i.e.}~along the whole geodesic.
The following results, in contrast, describe only the leading behaviour
in the near-singularity limit.

\vskip0.2cm
\noindent(iii)~~Decelerating universe, $0<\c<1$, for all $\k$:

Here, the first term in the denominator of \eqref{in} dominates
for small $a$, and the dependence of $a(u)$ on $u$ is exactly as in
case(i). So again we find, for all $\k$,
\EQ{
h_{ij} ~=~ {\c\over(\c+1)^2} {1\over u^2}\d_{ij} \ ,
\label{iq}
}
in the near-singularity limit. Notice that for the case of a matter dominated universe, $\c = \tfrac23 ~(w=0)$, the coefficient in \eqref{iq}
is $\tfrac 6{25}$, precisely the same as for the near-singularity limit 
of Schwarzschild, or Kerr, spacetime \cite{Blau2}.

\vskip0.2cm
\noindent(iv)~~Accelerating universe, $\c > 1$:

For an accelerating, $\c>1$, universe, we have already found the exact
solution \eqref{io} for $\k=0$.  For $\k=1$, the Friedmann equation does
not have a solution if $a$ becomes too small -- the parameters
$\k = 1, \c > 1$ do not correspond to a spacetime with an initial
singularity. This leaves the open universe $\k = -1$. This time, for
sufficiently small $a$, the second term in the denominator of
\eqref{in} dominates and we find $u \simeq \tfrac12a^2$, leading to
\EQ{
h_{ij} ~=~\delta_{ij} 
2^{-(1+{1\over\c})}\c^{-1} C_\c {1\over u^{1+{1\over\c}}} \ .
\label{ir}
}
The singular behaviour of $h_{ij}$ is softer than $1/u^2$ in this case;
this can be interpreted as due to a cancellation of the leading $1/u^2$
singular behaviour from the two terms in \eqref{if} when
$u = {1\over2}a^2$.

\vskip0.3cm

In order to consider the refractive index in these cosmological
examples, the following conceptual issue arises. Notice that the
expression for the index in \eqref{dm} and \eqref{dn}, 
depends on the past history of
the geodesic. The intuitive understanding is that a definition of the
refractive index involves waves coming in from past null infinity. 
In an FRW universe, the singularity cuts off this past
history and the refractive index cannot, as such, be defined in this
way. However,
another way to think about the vacuum polarization is in terms of an
initial value problem: rather than have the waves come in from past
null infinity, one defines a wave at some initial time. 
In this case one should think in terms of the retarded
Green function rather than the refractive index. This quantity was
considered in \cite{Hollowood:2008kq}. 
Effectively defining an initial value problem
provides an the upper limit for the 
$t$ integral in \eqref{dn} of the form $u_0$, where the 
initial value surface is at $u=u_0$:
\EQ{
\BcalF(u;z)=
\int_0^{u_0-i\epsilon}\frac{dt}{t^2}\,ie^{-izt}\left[
\BDelta\big(u,u-t\big)\sqrt{\det\BDelta\big(u,u-t\big)}-\Bone
\right]\ ,
\label{dn2}
}
The resulting quantity $\Bn(u;\omega)$ then plays the r\^ole of an
effective refractive index for waves that begin on the initial value
surface which, therefore, depends implicitly on the choice of
the initial value surface. 

The presence of the cut-off on the $t$ integral changes the low
frequency behaviour of the refractive index from the expressions in
\eqref{bo} and \eqref{bp} to
\EQ{
n_{ij}^\text{scalar}(u,u_0;\omega)=\delta_{ij}-\frac{\alpha}{360\pi m^2}
\big(R_{uu}(u)\d_{ij}+2R_{uiuj}(u)\big)F\Big(\frac{\omega}{m^2(u-u_0)}\Big)+
\cdots\ ,
\label{bo2}
}
while for spinor QED,
\EQ{
n_{ij}^\text{spinor}(u,u_0;\omega)=\delta_{ij}-\frac{\alpha}{180\pi m^2}
\big(13R_{uu}(u)\d_{ij}-4R_{uiuj}(u)\big)
F\Big(\frac{\omega}{m^2(u-u_0)}\Big)+\cdots\ ,
\label{bp2}
}
where 
\EQ{
F(x)=1-6\int_0^1 d\xi\,\xi(1-\xi)e^{-i/(2\xi(1-\xi)x)}\ .
}
The effect is that the low-frequency behaviour oscillates as a
function of $\omega$.

\begin{figure}[ht] 
\centerline{(a)\includegraphics[width=2.5in]{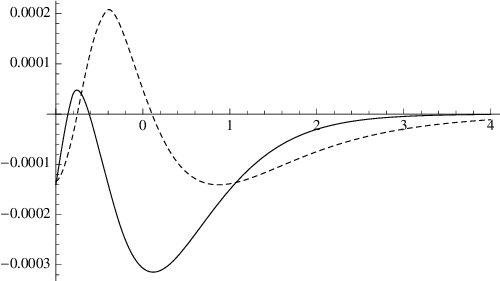}
\hspace{1cm}(b)\includegraphics[width=2.5in]{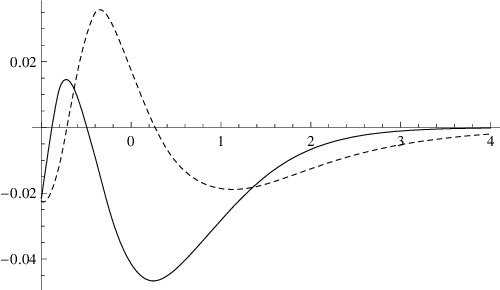}}
\caption{\footnotesize $\text{Re}\,n -1$ (continuous) and  
  $\text{Im}\,n$ (dashed)
  plotted as a function of $\log\omega$ in suitable units
  for the Milne universe for (a) scalar QED and (b) spinor QED.}
\label{f12}
\end{figure}

As an example of the kind of behaviour one encounters with
this definition of the refractive index, consider the Milne universe 
\cite{BD} which has spatially flat sections $\kappa=0$ and metric
\EQ{
ds^2=-dt^2+a^2t^2(dx^2+dy^2+dz^2)\ .
}
A congruence of null geodesics is simply obtained by taking $y$ and
$z$ constant along with $\dot t=a t\dot x$, which is solved by
taking $x=(2a)^{-1}\log u-x_0$, for constant $x_0$. 
The variables $(t,x_0,y,z)$ form a set of Rosen coordinates for
the congruence. Writing the metric in terms of these coordinates we
have
\EQ{
ds^2=-2a t\,dt\,dx_0+a^2t^2(dx_0^2+dy^2+dz^2)\ .
}
In order to take the Penrose limit, we choose an affine parameter
$u=at^2/2$ and identify $x_0=V$, $Y^1=y$ and $Y^2=z$. Then the
Penrose is easily taken giving, in both Rosen and Brinkmann coordinates,
\SP{
d\hat s^2&=-2du\,dV+2au\,\big(dY^1\,dY^1+dY^2\,dY^2\big)\\
&=-2du\,dv-\frac1{4u^2}\big((y^1)^2+(y^2)^2\big)du^2+dy^i\,dy^i\ .
}
This is precisely a singular homogeneous plane wave of the type
discussed in Section \ref{shpw} 
with $\alpha_i=0$ and the VVM matrix may be extracted from
\eqref{ffc} by taking the limit $\alpha_i\to0$:
\EQ{
\Delta_{ij}(u,u')=\delta_{ij}\frac{u-u'}{\sqrt{uu'}\log(u/u')}\ .
}
Numerical evaluations of the refractive index are shown in figure
\ref{f12} for scalar and spinor QED. 
In this case the spacetime is
conformally flat and so there is no bi-refringence. At low frequency
we see the characteristic oscillations that match \eqref{bo2} and \eqref{bp2}.

\vskip1cm

\section{Gravitational Waves}

Finally, we consider another major class of spacetimes -- gravitational
waves. Two examples are studied in detail, {\it viz.}~weak gravitational
waves and gravitational shockwaves, the latter being of particular
interest because of the use of the Aichelburg-Sexl metric 
\cite{Aichelburg:1970dh} 
in investigations of Planck energy scattering in quantum gravity.

\subsection{Weak gravitational waves}

The spacetime metric for a weak gravitational wave has the following
form in Rosen coordinates:
\EQ{
ds^2 ~=~ -2du\,dV + (1+\epsilon \cos\nu u)dY^{1}\,dY^1 
+ \big(1-\epsilon\cos\nu u\big)dY^2\,dY^2\ .
\label{ha}
}
Here, and in the following, $\epsilon$ is small and we work to linear
order. The transformation to Brinkmann coordinates is made using the 
zweibein
\EQ{
E_a{}^j(u)=\delta_a^i\big(1-(-1)^j\frac\epsilon2\cos\nu u\big)\ ,
\label{hb}
}
giving
\EQ{
ds^2 ~=~ - 2du\,dv - \frac{\epsilon\nu^2}2\cos\nu u\big[ 
(y^1)^2-(y^2)^2\big]\,du^2 + dy^j\,dy^j\ .
\label{hc}
}
Since $h_{11} = - h_{22} = {1\over2}\e \n^2 \cos\n u$, this is a 
Ricci flat spacetime and, as usual, will display gravitational birefringence. 

The metric \eqref{hc} is already in the form of the Penrose limit 
for a photon travelling in the opposite direction to the gravitational
wave along the null geodesic given by \eqref{ccc}. Notice that a
photon travelling in the same direction as the gravitational wave has
a flat Penrose limit and there is no vacuum polarization effect on its
refractive index \cite{Drummond:1979pp,Shore:2000bs}.

The Jacobi fields therefore satisfy
\EQ{
\frac{d^2y^{(\pm)}}{du^2}\pm\frac{\epsilon\nu^2}2\cos\nu u\, 
y^{(\pm)}=0\ ,
\label{hd}
}
where we have labelled the two polarizations 
$y^{(+)}\equiv y^1$ and $y^{(-)}\equiv y^2$.
This can easily be solved perturbatively in $\epsilon$,
with solution to linear order
\EQ{
y^{(\pm)}(u)=c_1+uc_2\pm\frac{\epsilon}2(c_1+uc_2)\cos\nu u
\mp\frac{\epsilon c_2}\nu\sin\nu u\ .
\label{he}
} 
The (diagonal) matrix $\BA(u,u')$ is the 
solution of the Jacobi equation subject to the now-familiar boundary
conditions and we find its components for the tow polarizations to be
\SP{
A_{(\pm)}(u,u')~=~ &u-u'\\ &\pm
\frac{\epsilon(u-u')}2\big(\cos\nu u+\cos\nu
u'\big)\mp\frac{\epsilon}{\nu}\big(\sin\nu u-\sin\nu u'\big)\ .
\label{hf}
}
This determines the eigenvalues of the Van-Vleck Morette matrix:
\EQ{
\Delta_{(\pm)}(u,u') ~=~ 1\mp\frac{\epsilon}2(\cos\nu u+\cos\nu u')
\pm\frac{\epsilon}{\nu(u-u')}(\sin\nu u-\sin\nu u')\ .
\label{hg}
}
Notice that this is regular for all $u,u'$, reflecting the fact that
in this weak field (small $\e$) approximation there are no conjugate
points in the null geodesic congruence.

The refractive index for scalar QED is given as usual by \eqref{dj}
where for the 2 polarizations we have
\SP{
{\cal F}^\text{scalar}_{(\pm)}(u;z)&=\mp\epsilon
\int_0^{\infty-i\epsilon}\frac{dt}{t^2}\,ie^{-izt}\\ &~~~\times
\Big[\frac12(\cos\nu u+\cos\nu (u-t))
-\frac1{\nu t}(\sin\nu u-\sin\nu (u-t))\Big]\\
&=\pm\epsilon\big[f(z)\cos\nu u+g(z)\sin\nu u\big]\ ,
\label{hh}
}
with
\SP{
f(z)~&=~\frac{z}{4\nu}\big[2\nu(1+\log
z)+(z-\nu)\log(z-\nu)-(z+\nu)\log(z+\nu)\big]\ ,\\
g(z)~&=~\frac{i}{4\nu}\big[\nu^2+2z^2\log z
+z(\nu-z)\log(z-\nu)-z(z+\nu)\log(z+\nu)\big]\ .
\label{hi}
}
For spinor QED the only difference relative to scalar QED 
is an overall multiple of 4:\footnote{In fact this generalizes to any
  Ricci flat background: the spinor and scalar QED results to linear
  order in the curvature, but allowing any number of derivatives, only
  differ by a factor of 4.}
\EQ{
{\cal F}^\text{spinor}_{(\pm)}(u;z)=4{\cal
  F}^\text{scalar}_{(\pm)}(u;z)\ ,
}
and so we will continue to discuss the scalar case.

The functions $f(z),~g(z)$ have branch points at $0$, $\infty$ and 
$z=\pm\nu$ and this means that $n_{(\pm)}(u;\omega)$ will have branch points
at 0 $\pm\infty$ and $\pm 2m^2/\nu$. In particular, the branch points
at $\pm 2m^2/\nu$ are points of non-analyticity of the refractive
index. This non-analyticity manifests itself by the fact that
$\text{Im}\,f(z)$ and $\text{Re}\,g(z)$ are
zero for $z\in{\bf R}>\nu$, while for $z\in{\bf R}<\nu$,
\EQ{
\text{Im}\,f(z)=\text{Re}\,g(z)=\frac{\pi
  z(\nu-z)}{4\nu}\ .
\label{hj}
}

The low frequency expansion of the refractive index follows immediately
from \eqref{hh}, \eqref{hi} and we find:
\SP{
n^{(\pm)}(u;\omega) ~=~1 ~&\pm~\frac{\alpha\epsilon\nu^2}{m^2\pi}
\Big[\frac1{360}+\frac1{6300}\left(\frac{\omega\nu}{m^2}\right)^2+
+\cdots\Big]\cos \nu u\\
&\mp~ i \frac{\alpha\epsilon\nu^2}{m^2\pi}
\Big[\frac1{840}\left(\frac{\omega\nu}{m^2}\right)
+\frac1{10395}\left(\frac{\omega\nu}{m^2}\right)^3+
\cdots\Big]\sin\nu u\ ,
\label{hk}
}
while at high frequencies,
\EQ{
n^{(\pm)}(u;\omega)=1\pm i\frac{\alpha\epsilon\nu}{6\pi\omega}
\sin\nu u+\cdots 
\label{hl}
}
Notice that the low frequency expansion is not sensitive to the
non-perturbative contribution \eqref{hj}. The leading terms for both
real and imaginary parts can however be read off from the effective
Lagrangian described in section 2.
\begin{figure}[ht] 
\centerline{(a)\includegraphics[width=2.5in]{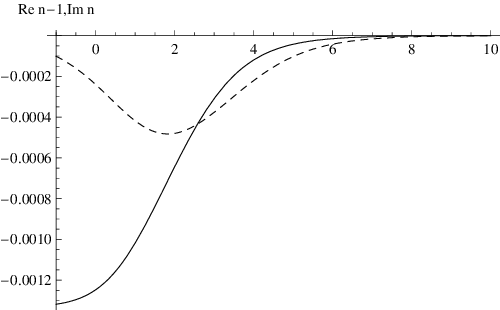}
\hspace{1cm}(b)\includegraphics[width=2.5in]{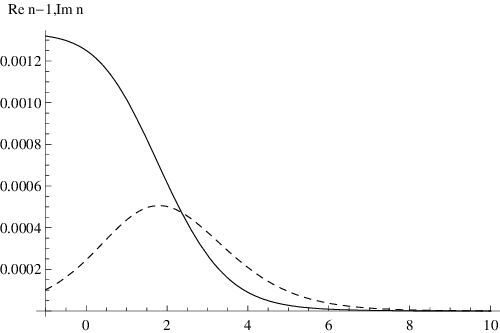}}
\caption{\footnotesize  $\text{Re}\,n -1$ and 
  $\text{Im}\,n$ for the 2 polarizations (a) and (b) for $u=0.2$, $m=\nu=1$
  plotted as a function of $\log\nu\omega/m^2$ in units of 
  $\epsilon\alpha$ for both polarizations. The point of non-analyticity 
  at $\omega=2m^2/\nu$ is quite clear. (Note that the fact that 
  the polarizations do not quite give mirror images is an artifact of 
  the numerical approximation.}  
\label{p16}
\end{figure}

The full form of the frequency dependence of the real and imaginary
parts of the refractive index is plotted numerically in Fig.~\ref{p16},
evaluated at a fixed point on the photon trajectory, for both scalar
and spinor QED. For the first polarization, has a $\RE n(\omega)$
which qualitatively looks similar to a conventional dielectric
with a single characteristic
frequency $\omega\sim m^2/\nu$, together with an imaginary 
part $\IM n(\omega)$ which is negative in the perturbative
small $\w$ regime before being dominated by the positive 
non-perturbative contribution \eqref{hj} above the branch point at 
$\w = 2m^2/\n$. Since this is a Ricci flat spacetime, the second 
polarization has the opposite sign, being superluminal at low
frequencies and with $\IM n(\omega)<0$ above the branch point,
normally an indication of amplification rather than dispersive 
scattering. The roles of the two polarizations of course change along 
the photon trajectory through the background gravitational wave.

\subsection{Gravitational shockwaves}

A gravitational shockwave is described by the Aichelburg-Sexl metric
\cite{Aichelburg:1970dh}
\EQ{
ds^2=-2du\,dv+f(r)\delta(u)\,du^2+dx^2+dy^2\ ,
\label{hha}
}
where $x=r\cos\phi$ and $y=r\sin\phi$. This describes an axis symmetric, 
plane-fronted wave moving at the speed of light, with a profile function
$f(r)$ in the transverse direction determined by the nature of the source:
a matter source $T_{uu} = \r(r)\d(u)$ corresponds to a profile function
satisfying $\D f(r) = -16\pi\r(r)$, where $\D$ is the 2-dim Laplacian.
We consider two sources of special interest: a particle with 
$\r(r) = \m \d(\underline{x})$, which gives a profile function
$f(r) = -4\m \log r^2$, and a homogeneous beam $\r(r) = \m$, for
which $f(r) = -4\pi\m r^2$.

The AS metric has many applications in general relativity and more
recently in quantum gravity and string theory, where it has been widely 
used in investigations of Planck energy scattering 
(see refs.\cite{ACV,GM,Veneziano:2004er} for a selection of papers). 
Recently, string scattering in a classical AS metric has been studied
\cite{Giddings:2007bw}
as a model of Planck scattering, and it would be interesting to
compare these results, characterized by the string scale, with
the equivalent QFT scattering, where the scale is set by the 
Compton wavelength of the particle in the vacuum polarization loop. 
For now, we focus on the effect of a shockwave collision on the 
refractive index in QED.

The null geodesics describing a photon travelling in the
opposite direction to the gravitational shockwave are well-known;
those corresponding to a homogeneous beam were first discussed 
by Ferrari, Pendenza and Veneziano in ref.\cite{Ferrari:1988cc}
(see also \cite{Shore:2002in}). 
With $u$ as the affine parameter, they satisfy
\SP{
r&=R+\tfrac12 f'(R)u\theta(u)\ ,\\
v&=V+\tfrac12f(R)\theta(u)+\tfrac18f'(R)^2u\theta(u)\ ,\\
\phi&=\Phi\ ,
\label{hhb}
}
where $(V,R,\Phi)$ are constants which label a congruence
of null geodesics. Consequently, these variables are a natural set of
Rosen coordinates. It is now straightforward to change variables from
$(u,v,r,\phi)$ to $(u,V,R,\Phi)$:
\SP{
dr&=dR+\tfrac12 f''(R)u\theta(u)dR+\tfrac12 f'(R)\theta(u)du ,\\
dv&=dV+\tfrac12f'(R)\theta(u)dR+\tfrac12f(R)\delta(u)du\\ &\qquad+\tfrac14
f'(R)f''(R)u\theta(u)dR+
\tfrac18f'(R)^2\theta(u)du\ ,\\
d\phi&=d\Phi\ ,
\label{hhc}
}
where we used the consistent replacement $u\delta(u)\to0$. The metric
in Rosen coordinates is therefore
\EQ{
ds^2=-2du\,dV+\Big(1+\tfrac12f''(R)u\theta(u)\Big)^2dR^2
+\Big(R+\tfrac12f'(R)u\theta(u)\Big)^2d\Phi^2\ .
\label{hhd}
}

We can now calculate the Penrose limit around a geodesic with $R=R_0$
by defining shifted Rosen coordinates
\EQ{
Y^1=R-R_0 ,\qquad Y^2=R_0\Phi\ ,
\label{hhe}
}
The metric is now in standard form and it is trivial
to implement the Penrose limit; we find
\EQ{
d\hat s^2=-2du\,dV+\Big(1+\tfrac12f''(R_0)u\theta(u)\Big)^2
(dY^1)^2+\Big(1+\tfrac12R_0^{-1}f'(R_0)u\theta(u)\Big)^2(dY^2)^2\ .
\label{hhf}
}
The zweibein $E^i{}_a(u)$ can be read off directly 
(see \eqref{cz}), and has non-vanishing components 
\EQ{
E^1{}_1(u)=1+\tfrac12f''(R_0)u\theta(u)\ ,
\qquad E^2{}_2(u)=1+\tfrac12R_0^{-1}f'(R_0)u\theta(u)\ .
\label{hhg}
}
We therefore find the Penrose limit in Brinkmann coordinates,
with a diagonal profile function $h_{ij}$ given by \eqref{ccd} as
\EQ{
h_{11}(u)=-\tfrac12f''(R_0)\delta(u)\ ,\qquad
h_{22}(u)=-\tfrac12R_0^{-1}f'(R_0)\delta(u)\ .
\label{hhh}
}
For the ``particle'' AS metric with $f(r)=-4\mu\log r^2$ we have
\EQ{
h_{ij}(u)=(-1)^i\delta_{ij}\frac{4\mu}{R_0^2}\delta(u)\ ,
\label{hhi}
}
which is an example of a Ricci flat plane wave. The related
``beam'' AS metric with $f(r)=-\mu r^2$ gives
\EQ{
h_{ij}(u)=\delta_{ij}\mu\delta(u)\ ,
\label{hhj}
}
which is a conformally flat shockwave.  

The Van Vleck-Morette matrix is found as usual by solving the
Jacobi equations subject to the appropriate boundary conditions,
with $h_{ij}$ given by \eqref{hhi}, \eqref{hhj}.
Writing the above zweibeins as 
$E^i{}_a = \bigl(1 -\l_i u \theta(u)\bigr)\d^i{}_a$,~
so that $h_{ij} = \l_i \d_{ij}$,~ we find, in general, if $u$ and $u'$
are on different sides of the shockwave, $uu'<0$,
\EQ{
\Delta_{ij}(u,u') ~=~ \frac{|u-u'|}{|u-u'|+\lambda_i u u'} \d_{ij}\ , 
\label{hhk}
}
while $\D_{ij}(u,u') = 1$ ~if $u,u'$ are on the same side of the 
shockwave, $uu'>0$.
When $uu'<0$ for a polarization
with $\l_i > 0$, there are conjugate points with  
$|u| +|u'| = \l_i |u u'|$ 
provided the gravitational field of the 
shockwave is sufficiently strong, $\l_i |u| > 1$ and $\l_i |u'| > 1$. In fact,
the relation for conjugate points is
\EQ{
\frac1{|u|}+\frac1{|u'|}=\lambda_i\ ,
}
which is the lens formula for a lens of focal length $\lambda_i^{-1}$. 
When the $\lambda_i>0$ the lens is
converging and a conjugate point exists as long as $u$ and $u'$ are large
enough, as above. In this case, the refractive index integral will
have a branch point on the real $t$ axis. 
Correspondingly, when $\lambda_i<0$ the lens is
diverging and no conjugate points exist.

The refractive index is given by the formulae in sections 4 ad 5
and is easily evaluated for both the particle (Ricci-flat) and
beam (conformally flat) AS spacetimes with either scalar or spinor QED. 
For example, consider the conformally flat shockwave and scalar QED. 
We have, for either polarization,
\SP{
{\cal F}(u;z)~&=~ 
\theta(u)\int_u^{\infty-i\epsilon}\frac{dt}{t^2}\,ie^{-izt}\,
\left[\Big(\frac t{t+\lambda u(u-t)}\Big)^2-1\right]\ ,\\
&=~ \theta(u)\Big[z\Gamma(-1,iuz)-\frac{z}{(1-u\lambda)^2}
e^{\tfrac{i\lambda u^2z}{1-\lambda u}}\Gamma\big(-1,\tfrac{iuz}
{1-\lambda u}\big)\Big]
\label{hhl}
}
where $\Gamma(a,z)=\int_z^\infty t^{a-1}e^{-t}\,dt$ 
is the incomplete gamma function. Notice that the lower limit of 
the integral is at $t=u$, in order that the point $u'=u-t < 0$ is on 
the other side of the shockwave. The factor of $\theta(u)$ ensures
that the shockwave lies in the past of the point $u$.

\begin{figure}[ht] 
\centerline{\includegraphics[width=2.5in]{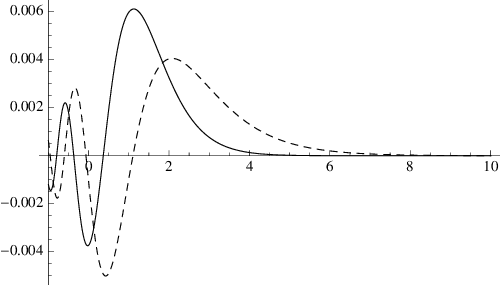}
\hspace{0.2cm}\includegraphics[width=2.5in]{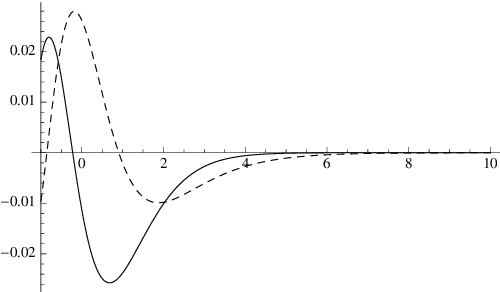}}
\caption{\footnotesize  $n-1$ (real part continuous, imaginary part
  dashed) for $\lambda=1$ and $u=2$ (left) and $u=\tfrac12$ (right) 
  plotted as a function of $\log\omega$ for scalar QED.}
\label{f7}
\end{figure}

In this case, $\l$ is negative and the integrand of ${\cal F}(u;z)$ 
has no poles or branch points on the real $t$ axis. Nevertheless,
inspection of the explicit solution in terms of gamma functions 
makes it clear that ${\cal F}(u;z)$ itself still has branch cuts
from $z=0$ to $\infty$. 
The refractive index $n(\w)$ evaluated numerically from \eqref{hhl}
is shown in Fig.~\ref{f7}. 

The small $\omega$ behaviour of the refractive index can be obtained 
analytically by expanding the integrand of ${\cal F}_+(u;z)$ in powers 
of $t-u$. In all cases, 
\SP{
{\cal F}(u;z)~&=~ \theta(u)\int_u^{\infty-i\epsilon}dt\,ie^{-izt}\,
\Big[-\lambda a_0(t-u)^\gamma/u^{\gamma+1}+{\cal O}(t-u)^2\Big]\\ &=
-i\theta(u)a_0e^{-izu}(izu)^{-\gamma-1}\Gamma(1+\gamma)+\cdots\ ,
\label{hhm}
}
where, depending on the case, 
$\gamma=0$ or $1$ and $a_0$ is some $\lambda u$ dependent constant.
The $\xi$ integral is then dominated by a stationary phase with
$\xi=\tfrac12$ which gives a behaviour
\EQ{
n(u;\omega)~\underset{\omega\to0}{=}~
1+\alpha C(u,m,\lambda)\theta(u)\omega^{\gamma+1/2} 
e^{-2im^2 u/\omega}\ ,
\label{hhn}
}
where $C(u,m,\lambda)$ is some $\omega$ independent factor. 
This shows clearly how in this case the non-vanishing imaginary
part of the refractive index arises from the
oscillating phase factor in the integrand evaluated at the
lower limit of the $t$-integral, even though there are no 
singularities or branch points off the real axis.
We can make these formulae even more explicit for the conformally 
flat scalar QED case. In this case from \eqref{hhl}, we get
\EQ{
n(u;\omega) ~\underset{\omega\to0}{=}~
1+\alpha\lambda\sqrt{\frac{i\pi\omega^{3}}{2^{11}u^{5}m^{10}}}\theta(u)
e^{-2ium^2/\omega}\ .
\label{hho}
}
On the other hand, for large $\omega$,
\EQ{
n(u;\omega) ~\underset{\omega\to\infty}{=}~
1+\frac{i\alpha\lambda}{12(u\lambda-1)\omega}\theta(u)+\log(c\,\omega)\,{\cal
  O}\Big(\frac1{\omega^2}\Big)\ . 
\label{hhp}
}
Note the oscillations at small $\omega$ as shown in the numerical
evaluation in Fig.~\ref{f7}.

It is worth pointing out that in this case we cannot match the low
frequency behaviour to the effective action because the curvature is
not differentiable. The other thing this computation makes clear is
the non-local natural of vacuum polarization: the refractive index 
is non-trivial even in a region of flat space because
the $e^+e^-$ loop can extend far enough back in time to touch 
the shockwave.

\section{Discussion}

The propagation of light, and other quantum fields, is a fundamental
and difficult problem for QFT in curved spacetime. In particular,
it has raised the question of how the original Drummond and Hathrell
discovery of low-frequency superluminal propagation could be consistent
with causality. In our previous work, we have resolved this issue
by developing techniques which allow an explicit computation of the
full frequency dependence of the refractive index and Green functions
for (scalar) QED in curved spacetime.

In the present paper, we have further developed the ``phenomenology'' of
the refractive index in a number of important spacetimes, namely black
holes, gravitational waves and FRW universes, and extended our analysis
to spinor QED. A crucial observation is that, to leading order in
$R/m^2$, the refractive index depends only on the Penrose limit of the
background spacetime around the original null geodesic. This is a very
powerful simplification and reveals a number of universal features. For
example, propagation in many examples reduces in the Penrose limit
to the simple case of homogeneous plane waves, especially the singular
homogeneous plane waves which are associated with the near-singularity
limits for both black holes and cosmological spacetimes.

A key element in understanding the causal properties of light
propagation, and more generally the retarded and advanced Green
functions, is the analytic structure of the refractive index.
As we have shown, this is in turn controlled by the Van Vleck-Morette
matrix which describes the geometry of geodesic deviation. The examples
we have studied here display a rich variety of analytic structures,
very different from the simple picture in flat spacetime. The Green
functions and refractive index therefore display many interesting
and physically important features special to curved spacetimes; in
particular, conventional dispersion relation theory in QFT,
exemplified by the Kramers-Kronig relation, is radically changed.
Nevertheless, in all cases we have investigated, the analytic
structure still allows the definition of retarded and advanced
Green functions with the required causal properties. We can conclude
that QED in curved spacetime is indeed a perfectly causal theory
despite the unusual and remarkable occurrence of superluminal phase
velocities at low frequencies.

Whilst this goal has been achieved and one can finally see explicitly
how a low-frequency phase velocity bigger than $c$ is compatible with
retarded Green functions that vanish outside the lightcone, a new
puzzle arises in that the imaginary part of the refractive index
is found in many examples to be negative. This indicates that the
optical theorem is violated in curved spacetime. A full explanation
of this phenomenon will be presented elsewhere \cite{Optical}.

\begin{center}
{\tiny **********************************}
\end{center}

TJH and GMS acknowledge the support of STFC grant ST/G000506/1.
RS would like to acknowledge the support of STFC studentship ST/F00706X/1.

\end{document}